%% file: main.tex
\documentclass[sigconf,10pt]{acmart}
\renewcommand\footnotetextcopyrightpermission[1]{} 
\pdfoutput=1
\setcopyright{none}
\usepackage{booktabs} 
\usepackage{subfigure}
\usepackage{graphicx}
\usepackage{comment}
\usepackage{xspace}
\usepackage{subfigure}
\usepackage{url}
\usepackage{color}
\usepackage{multirow}
\usepackage{mdwlist}
\usepackage{amsmath}
\usepackage{amssymb}
\usepackage[latin1]{inputenc}
\usepackage[T1]{fontenc}
\usepackage{mwe}

\usepackage{enumitem}
\usepackage[lined,boxed,commentsnumbered]{algorithm2e}
\usepackage{flexisym}
\usepackage{siunitx}
\newcommand{\fakeparagraph}[1]{\vspace{.5mm}\noindent\textbf{#1.}}
\newcommand{\fakepar}[1]{\fakeparagraph{#1}}

\newcommand{\systemnosc}{TunnelScatter\xspace}
\usepackage{wrapfig}
\usepackage{adjustbox}

\usepackage{enumitem}
\usepackage{amssymb}


\title{TunnelScatter: Low Power Communication for Sensor Tags using Tunnel Diodes}
\author{Ambuj Varshney}
\affiliation{%
  \institution{Uppsala University, Sweden}
}
\email{ambuj.varshney@it.uu.se}

\author{Andreas Soleiman}
\affiliation{%
  \institution{Uppsala University, Sweden}
}
\email{andreas.soleiman@it.uu.se}

\author{Thiemo Voigt}
\affiliation{%
  \institution{Uppsala University, Sweden\\
  RISE SICS, Sweden}
}
\email{thiemo.voigt@it.uu.se}

\renewcommand{\shortauthors}{Varshney et al.}

\setcopyright{acmcopyright}
\begin{document}
\fancyhead{}

\begin{abstract} 
\input{abstract}

\end{abstract}

\maketitle

\section{Introduction}
\input{intro-new.tex}

\section{Related work}
\label{related-work}
\input{related.tex}

\section{System Design}

\input{sections/architecture}

We show an overview of TunnelTag in Figure~\ref{fig:tunneloverview}. In this section, we describe TunnelTag's components with a greater emphasis on the TunnelScatter mechanism. 

\input{sections/vls.tex}

\section{Evaluation}
\label{evaluation}
\input{sections/eval.tex}

\section{Conclusions}

We have presented TunnelScatter, a mechanism that uses a tunnel diode to enable low-power communication when the ACS is weak or even absent. We demonstrate that TunnelScatter enables communication in challenging environments even without ACS. Further, in the presence of a weak ACS, TunnelScatter supports multi-floor communication, a scenario where state-of-the-art backscatter systems achieve a range of only a few meters.  We employ TunnelScatter on TunnelTags, that can sense physical phenomena and transmit them at a peak biasing  power of \SI{57}{\micro\watt}. We demonstrate hand gesture sensing as a concrete use case of TunnelScatter.

\fakepar{Acknowledgements}
We thank our shepherd, Omid Abari, and the reviewers whose insightful comments and guidance have been very helpful in improving this paper. We  thank Luca Mottola, Elena Di Lascio and Christian Rohner for their constructive feedback. We thank Domenico Giustiniano for suggesting the name TunnelTag. This work has been partly funded by the Swedish Research Council (VR, grant 2018-05480), VINNOVA, the Swedish Foundation for Strategic Research (SSF) and a Google Faculty Research Award. 

\bibliographystyle{ACM-Reference-Format} 
\bibliography{ref}

\end{document}

%% file: abstract.tex
Due to extremely low power consumption, backscatter has become the transmission mechanism of choice for battery-free devices that operate on harvested energy. However, a limitation of recent backscatter systems is that the communication range scales with the strength of the ambient carrier signal~(ACS). This means that to achieve a long range, a backscatter tag needs to reflect a strong ACS, which in practice means that it needs to be close to an ACS emitter. We present \emph{TunnelScatter}, a mechanism that overcomes this limitation. TunnelScatter uses a tunnel diode-based radio frequency oscillator to enable transmissions when there is no ACS, and the same oscillator as a reflection amplifier to support backscatter transmissions when the ACS is weak. Our results show that even without an ACS, TunnelScatter is able to transmit through several walls covering a distance of \SI{18}{\meter} while consuming a peak biasing power of \SI{57}{\micro\watt}. Based on TunnelScatter, we design battery-free sensor tags, called \emph{TunnelTags}, that can sense physical phenomena and transmit them using the TunnelScatter mechanism.

%% file: intro-new.tex
\label{sec:intro}

\begin{figure}[!tb]
\centering
\subfigure[Amplified Backscatter Transmissions~(ABT) \label{carrierbased}]{\includegraphics[width=0.78\linewidth]{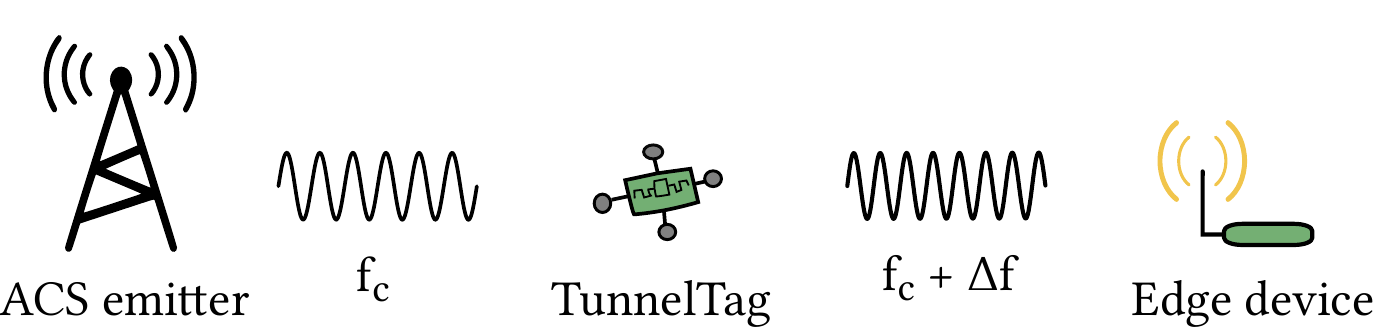}}\quad
\subfigure[Ambient Carrier-less Transmissions~(ACLT) \label{carrierless}]{ \includegraphics[width=0.96\linewidth]{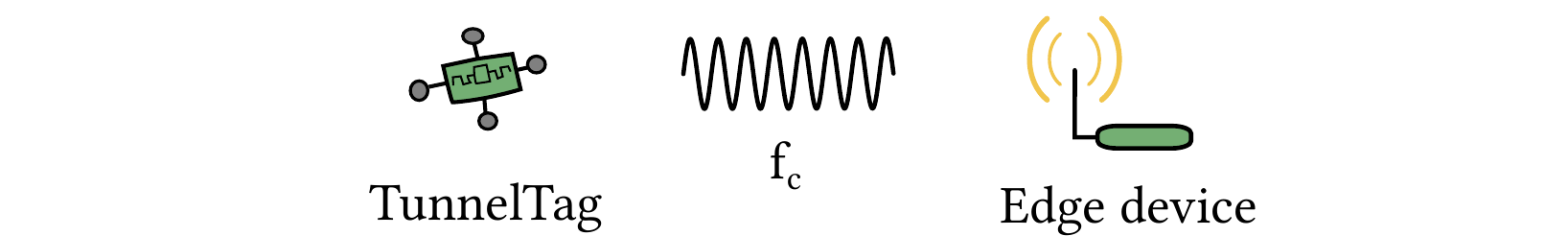}}\quad
\vspace{-5mm}
\caption{TunnelTags sense physical phenomena and perform transmissions using the tunnel-diode based TunnelScatter mechanism, which operates at a peak biasing power of \SI{57}{\micro\watt}. TunnelScatter achieves this task by backscattering an ambient carrier signal~(ACS) (a), or in the absence of an ACS, by locally generating and modulating a signal with sensor readings (b).}  
\vspace{-5mm}
\label{fig:systemoverview}
\end{figure}

Recent years have seen a rapid growth of sensing applications. These applications are commonly enabled through battery-powered devices. However, large-scale deployments with such devices suffer from high cost (tens to hundreds of USD per device), overhead of maintenance for replacing depleted batteries, and deployment inconvenience due to bulky form factor of devices. Thus, battery-powered sensor devices present a significant challenge to 
the vision of ubiquitous sensing~\cite{sustplanet}. As a result, there has been a growing interest in battery-free \emph{sensor tags}~\cite{batteryfreehdcamera,batteryfreecellphone,bfvls,lorabackscatter, witag,interscatter}. 

Sensor tags eliminate batteries by operating on energy that is harvested from ambient sources such as light~\cite{bfvls} or wireless signals~\cite{powifi,naderiparizi2015wispcam}. 
It is, however, challenging to transmit on minuscule and intermittent energy using conventional wireless transmission mechanisms~\cite{passivewifi,interscatter,lorea,lorabackscatter}.

Backscatter communication overcomes this limitation by enabling wireless transmissions at tens of \SI{}{\micro\watt}s of power~\cite{lorea,passivewifi,lorabackscatter,interscatter,lorabackscatter}. Consequently, backscatter has emerged as the mechanism of choice to enable battery-free sensor tags. 

\fakepar{Challenge} Backscatter enables wireless transmissions by reflecting or absorbing ambient wireless signals~\cite{ambientbackscatter,passivewifi,griffin2009complete}. This process is realized by changing the impedance of the antenna, an operation that can be performed at sub-\SI{}{\micro\watt} of power~\cite{ambientbackscatter} which leads to very low power consumption of backscatter systems. Backscatter as a transmission mechanism is not new; one of the first systems was demonstrated more than half a century ago~\cite{sealbug}. Backscatter mechanisms have also been used for decades in RFID systems~\cite{want2006introduction}. Although the underlying concept of backscatter communication has been known for decades, it is recently that backscatter systems have made significant progress. Recent systems are able to generate commodity wireless transmissions such as WiFi~\cite{passivewifi}, ZigBee~\cite{interscatter}, BLE~\cite{ensworth2015every} and LoRa~\cite{lorabackscatter}, or achieve a large~(\SI{}{\kilo\meter}) communication range~\cite{lorea,lorabackscatter,peng2018plora}. These advances in backscatter systems have also enabled novel applications~\cite{lorabackscatter,interscatter,batteryfreehdcamera,batteryfreecellphone,bfvls} such as battery-free cameras.~\cite{batteryfreehdcamera,naderiparizi2015wispcam,backcam}, audio transmissions~\cite{batteryfreecellphone,talla2013hybrid}, and activity and gesture recognition~\cite{barnet,bfvls}.

Despite this success, limitations that  hinder practical deployments remain;
backscatter systems require the presence of a strong  strong ambient carrier signal~(ACS). This requires the tag to be located in close proximity~(\SI{1}{\meter}) of a strong ACS emitter~($>$ \SI{500}{\milli\watt}) to achieve the highest and often practical range~\cite{passivewifi,lorea,peng2018plora}. Even backscatter systems that leverage ambient signals instead of a dedicated ACS emitter encounter this problem.  Ambient backscatter systems achieve a range that is sufficient to enable many applications when 
the backscatter tag is in  proximity to a RF signal source, i.e., near a TV~\cite{ambientbackscatter} or FM tower~\cite{fmbackscatter}, or located near a WiFi or any other wireless device~\cite{freqshift,wifibackscatter,freerider,hitchhike}.
Furthermore, the strong ACS may also cause co-existence issues for other wireless devices that are sharing the spectrum~\cite{backvlc}. 
 
\fakepar{Contributions} Backscatter systems require the presence of a strong ACS at the tag, which usually limits the application scenarios. To significantly improve the communication range of backscatter systems when the ACS is weak or even when it is absent,
we present the \emph{TunnelScatter} low power communication mechanism. 
TunnelScatter is designed using a semiconductor device tunnel diode which shows a region of negative resistance~(RNR)~\cite{rca1963rca}. This RNR enables the use of tunnel diodes in a variety of RF applications, for example, as oscillators~\cite{rca1963rca}, amplifiers~\cite{amato2018tunneling,rca1963rca}, or switches~\cite{rca1963rca}, and as we show in this paper, to enhance backscatter systems.

To support operation without an ACS, we design the  ambient carrier-less transmissions~(ACLT) mode of the  TunnelScatter mechanism. This mode uses a tunnel diode as an RF oscillator to generate a signal at a frequency band of \SI{868}{\mega\hertz}. This signal is then modulated using amplitude shift keying~(ASK) to encode sensor readings. To perform this operation, we have to maintain the tunnel diode in RNR, for which we consume a peak biasing power of of \SI{57}{\micro\watt}. The ACLT mode enables  sensor tags to communicate without the ACS, which existing backscatter systems require.

TunnelScatter operates with enhanced capabilities in the presence of an ACS. We observe that in the presence of an ACS, the tunnel diode oscillator~(TDO) latches onto the ACS through a process called \emph{injection locking}~\cite{injectionlocking,rca1963rca} and behaves as a reflection amplifier~\cite{amato2018tunneling}. This results in a significant gain while backscattering.  Based upon this concept, we design a long-range mode we call amplified backscatter transmissions~(ABT). This mode encodes sensor readings using frequency shift keying~(FSK), and
shows orders of magnitude improvement in communication range as compared to LoRea~\cite{lorea}, when backscattering a weak ACS. However, when the ACS is strong, we observe that the reflection amplifier performs poorly compared to a conventional backscatter tag.  
Hence, we also design a mechanism that senses the ACS' strength and uses the tunnel diode only when the ACS is weak. This ensures that TunnelScatter exploits the full range of the ACS strengths that an application may encounter.

\fakepar{Battery-free Tags and Applications} We employ Tunnel\-Scatter on battery-free sensor tags that we call \emph{TunnelTags}. As shown in Figure~\ref{fig:systemoverview}, TunnelTags can communicate even without an ACS. Therefore, the ACS emitter is an optional component to our tag. A complete system also includes an edge device to receive and process the transmitted signals. 

To sense physical phenomena, we design a sensing component that we call self-sustaining sensor. Many sensing phenomena can also be the source of energy. For example, light can be used for both sensing and harvesting~\cite{bfvls}. We design the self-sustaining sensor by coupling an energy harvester with a sensor. This supports battery-free operation due to low power consumption of sensors~\cite{ekhonet}.

Before sensor readings can be transmitted, they need to be digitised. The energy harvesting conditions can vary significantly during a deployment; as an example, light levels are much lower during the night than during the day. However, digitization using high resolution ADCs can be prohibitively energy expensive under poor energy harvesting conditions. Hence, we design a mechanism that we call polymorphic processing pipeline~(PPP) that adapts the digitization resolution and capabilities to the energy harvesting conditions. 

As we discuss in Section~\ref{appusecase}, our system enables several application scenarios. As an example, integrating self sustaining temperature sensors could enable TunnelTags to track the ambient room temperature, or accelerometers to enable infrastructure monitoring~\cite{brimon}. These applications fulfill the requirement that the harvester needs some time to charge the capacitors before the TunnelTag can operate. As a application use case of our system, we prototype and demonstrate a self-sustaining light sensor, which as we show in Section~\ref{appusecase}, enables hand gesture recognition.

\fakepar{Summary of results} We focus on the results obtained using TunnelScatter, the key component of the system.
\begin{itemize}[noitemsep,leftmargin=*]
  \item TunnelScatter, unlike backscatter, enables transmissions without requiring an ACS. It enables us to communicate through several walls covering a distance of \SI{18}{\meter}.
  \item When backscattering, TunnelScatter adapts to strength of the ACS: with a weak carrier signal, it backscatters with amplification achieving multi-floor communication. In comparison, a tag based on state-of-the-art LoRea~\cite{lorea} under similar conditions achieves a range of \SI{3}{\meter}. 
  \end{itemize}

The design of TunnelTag and TunnelScatter includes several hardware innovations. We devise a  digitisation mechanism that is inspired by wake-up radios and adapts its digitisation capabilities to the current energy harvesting conditions. We design a novel switchover mechanism that selects between the tunnel diode and the conventional RF switch for transmissions based on the ACS' strength. Finally, we also design a proof-of-concept self-sustaining light sensor designed to operate under diverse light conditions.

%% file: related.tex
We are not the first to use tunnel diodes to improve the  range of backscatter systems, and we build on existing works~\cite{amato2018tunneling,amato2018tunnel2,amato2017achieving,amato2015long}. However,  when compared to these existing works, we are the first to integrate a tunnel diode in a long-range backscatter system~\cite{lorea}, and demonstrate orders of magnitude improvement in  communication range when receiving with commodity radio transceivers~\cite{cc1310launchpad}. Also, we design a mechanism to enable operation of tunnel diode based tag  with ACS's of diverse strengths. Building on earlier works that use a tunnel diode as an oscillator~\cite{rca1963rca,4066401}, we demonstrate experimentally that tunnel diodes can function as an oscillator for sensor tags that can now communicate without an ACS.  Below, we discuss related work in detail.

\fakepar{Backscatter systems} There has been a  interest in backscatter communication. Recent systems show the ability to generate WiFi~\cite{passivewifi,interscatter}, Bluetooth~\cite{ensworth2015every}, ZigBee~\cite{interscatter,zigbeecarlos}, and LoRa~\cite{lorabackscatter,peng2018plora,3dloc} transmissions at tens of \SI{}{\micro\watt}s of power consumption. Other efforts transmit over large distances~(kilometres)~\cite{peng2018plora,lorea, lorabackscatter}. All of these systems need a strong ACS emitter and require that the tags are located in proximity to the ACS emitter to achieve the highest communication range. The constraint of proximity to a strong ACS emitter hinders potential application scenarios. In contrast,  \systemnosc enables a large range even when the ACS is weak, for example, when the tag is not close to the ACS emitter.   

\fakepar{Ambient backscatter} Our work is related to backscatter systems that reflect ambient signals, and do not require a dedicated ACS emitter. Ambient backscatter reflects TV signals~\cite{ambientbackscatter,turbocharging} and requires the vicinity of TV towers. FM backscatter overcomes this limitation by taking advantage of the ubiquitous nature of FM signals~\cite{fmbackscatter}. Analog FM is, however, being phased out which limits FM backscatter. Furthermore, it requires a large antenna which limits practical deployment for many applications. Other systems such as WiFi backscatter~\cite{wifibackscatter}, FreeRider~\cite{freerider}, HitchHike~\cite{hitchhike}, BackFi~\cite{backfi}, and PLoRa~\cite{peng2018plora} can achieve sufficient range only in proximity to the ambient signal source (WiFi router or LoRa node). We overcome these limitations; \systemnosc introduces a new modality that enables wireless transmissions for sensor tags without requiring any ambient signal.

\fakepar{Reflection amplifiers in backscatter systems} Kimionis et al. design a reflection amplifier  using a RF transistor with a gain as high as \SI{10.2}{\decibel} in the \SI{900}-\SI{930}{\mega\hertz} band at a power consumption of \SI{325}{\micro\watt}~\cite{transistoramplifier}. Amatao et al. develop a reflection amplifier using a tunnel diode that achieves a gain as high as \SI{40}{\decibel} in the \SI{5.8}{\giga\hertz} band while consuming \SI{45}{\micro\watt}s of power~\cite{amato2015long,amato2018tunneling}. Further, they demonstrate improvements in range using a signal analyser as a receiver. We build \systemnosc on Amatao et al.~\cite{amato2018tunneling,amato2017achieving,amato2015long,amato2018tunnel2} and design a reflection amplifier for the widely used  868\SI{}{\mega\hertz} band. We experimentally demonstrate that as the carrier signal strength increases, a conventional backscatter tag designed using RF switch starts to outperform the tunnel diode reflection amplifier. Therefore, we design a low-power switchover mechanism to select between the reflection amplifier and the conventional tag depending on the strength of the ambient carrier signal. We are the first to demonstrate improvements in range using tunnel diodes with commodity radio transceivers as receivers, which enables a ubiquitous deployment of backscatter readers~\cite{lorea,ensworth2015every,passivewifi}. Finally, we also experimentally demonstrate the ability of the tunnel diode oscillator~(TDO) to enable transmissions without requiring ACS for sensor tags.

\fakepar{Oscillators in backscatter systems} Recent systems use commercial precision oscillators to support frequency shift backscatter~\cite{lorea,bfvls,rfbandid}. The power consumption of these oscillators increases with frequency~\cite{ltc6906}. This becomes prohibitively energy-expensive for sensor tags at higher frequencies; for example, they consume~\SI{2}{\milli\watt} at a frequency of \SI{10}{\mega\hertz}~\cite{ltc6900}. Due to the high power consumption of precision oscillators, Abedi et al.~\cite{witag} design a mechanism which shifts backscatter signals at a much lower frequency of \SI{50}{\kilo\hertz}. Zhang et al.~\cite{freqshift} overcome the energy limitations of commercial oscillators by trading off accuracy for power consumption. They design a ring oscillator that oscillates at a frequency of \SI{20}{\mega\hertz} at \SI{21}{\micro\watt} of power, to enable frequency-shift backscatter. However, all of these designs are unsuitable to support ACLT  for sensor tags, as ACLT requires oscillators operating at \emph{hundreds of \SI{}{\mega\hertz}} at tens of \SI{}{\micro\watt}s of power consumption. TunnelScatter achieves this  using TDO. 

\fakepar{Processing overhead}  Zhang et al. show that performing local processing is significantly  energy expensive than backscatter transmissions~\cite{ekhonet}.  Talla et al. couple a microphone to a backscatter transmitter to transmit audio signals at \SI{3.48}{\micro\watt} of power~\cite{batteryfreecellphone}.  Naderiparizi et al.~\cite{batteryfreehdcamera} design an analog backscatter technique that transmits  signals from image sensors, and also frequency shift the backscattered transmission. We design a hand gesture sensing system that avoids computational elements~\cite{bfvls}. All of these systems are limited in that they either use a complicated and expensive SDR for reception~\cite{batteryfreehdcamera,batteryfreecellphone} which negatively impacts the prospect of pervasive backscatter readers~\cite{lorea}, or they are constrained by the sensing resolution and communication range ~\cite{bfvls,batteryfreehdcamera,batteryfreecellphone}. We advocate a balanced approach  that avoids computational elements under poor energy harvesting conditions but supports higher resolution when energy harvesting conditions permit. We also significantly improve our communication ability through TunnelScatter mechanism.

%% file: sections/architecture.tex
Our system is divided into three components: a generic battery-free sensor tag that we call TunnelTag that transmits using the TunnelScatter mechanism, an edge device comprising of one or more RF receivers to receive and interpret sensor readings, and an optional ACS emitter, as shown in Figure~\ref{fig:systemoverview}. As the TunnelTag, and in particular TunnelScatter are our key contributions, we discuss them in Section~4. 

At a high level, our system performs a series of steps as follows: First, the TunnelTag using a self-sustaining sensor harvests energy from an ambient source such as light and charges a supercapacitor.  TunnelTag uses this energy to support the operation of the tag. Next, when there is an event of interest such as a temperature change, or a change in the ambient light conditions, the polymorphic processing pipeline~(PPP) depending on the energy harvesting condition, processes and digitises the sensor readings at a high or a low resolution. Finally, the digitised sensor readings are transmitted using the \systemnosc mechanism, and received and interpreted by an edge device.  

\fakepar{Ambient carrier signal emitter} In the presence of an ACS, \systemnosc operates in ABT mode, and transmits sensor readings using backscatter communication. To generate the ambient carrier signal, we use a software-defined radio~(SDR), USRP B200~\cite{usrp}, as ACS emitter.  We use an SDR to generate the ACS, as it allows us to control the strength of the ACS at the tag. This, in turn, enables us to perform controlled experiments with weak ACS strengths where amplification using tunnel diodes is important. However, our system can also employ commodity and low-cost radio transceivers to generate the ACS, which enables commodity devices to act as ACS emitters. To support the use of commodity radio transceivers for ACS generation, we  can either build on LoRea~\cite{lorea} or Interscatter~\cite{interscatter}.

\fakepar{Edge device}  In our system, at the edge device, we do not use SDRs or RFID readers to receive backscatter transmissions, as is commonly done~\cite{want2006introduction}.  Instead, we leverage a low-cost radio transceiver~(< 10~USD~), the Texas Instruments CC1310 ~\cite{cc1310launchpad} with a high sensitivity~(\SI{-124}{\decibel}m), as a receiver. Reception using low cost transceivers enables ubiquitous deployments of backscatter readers~\cite{lorea}. The transceiver can be used to receive FSK backscatter transmissions when TunnelScatter operates in ABT mode, or to function as an energy detector to receive amplitude-modulated transmissions when TunnelScatter operates in ACLT mode. Hence, we equip the edge device with two CC1310 receivers, to receive transmissions from TunnelScatter operating in either ABT or ACLT mode.

\begin{figure}[t!]
    \centering
    \includegraphics[width=\columnwidth]{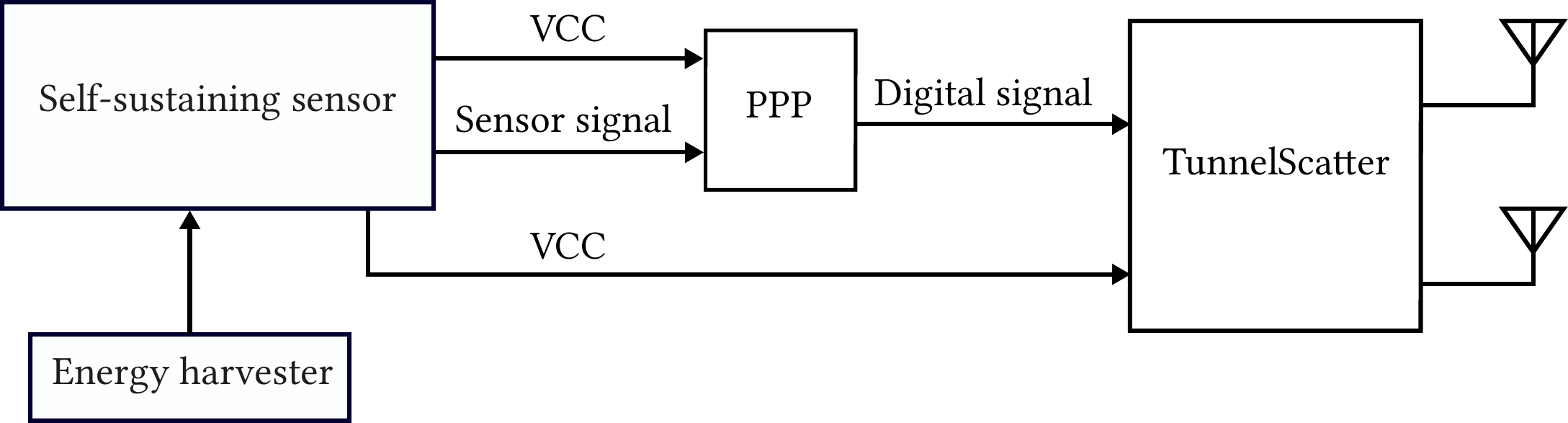}
    \vspace{-4mm}
    \caption{\emph{Overview of TunnelTag.} It is divided into three components; self-sustaining sensor, PPP, and TunnelScatter mechanism.}
        \vspace{-4mm}
    \label{fig:tunneloverview}
\end{figure} 
 To receive the amplitude-modulated transmissions,  we continuously gather the energy measurements by performing Received Signal Strength~(RSS) sampling at the frequency of the transmissions from the TunnelTag. We reconstruct the sensor readings from the received samples by first determining the average noise floor, and then by approximating all values above it to binary ones, and below it to binary zeroes. The bitrate in the ACLT mode is restricted by the RSS sampling rate of the CC1310 receiver, which we found in our implementation to be \SI{10}{\kilo\hertz} restricting bitrate to 1 kbps. We believe this is an engineering constraint, as FS-Backscatter~\cite{freqshift} demonstrates a bitrate of 50~kbps, using ASK and RSS sampling. On the other hand, to receive backscatter transmissions, we build on LoRea, and demonstrate a bitrate of 2.9 kbps in this paper, while we can also support bitrates as high as 100 kbps, as reported by LoRea~\cite{lorea}. Once the sensor readings are received, the edge device can process them according to the target application. 

\section{TunnelTag}

%% file: sections/vls.tex
 \begin{table}
\vspace{-2mm}
\caption{Example of low-power sensors that can be integrated with  self-sustaining sensor.}
\label{sensorself}
\begin{adjustbox}{width=0.7\columnwidth,center}  
\begin{tabular}{|l|l|l|}
\hline
Sensor Type   & Name    & Power                                                                   \\ \hline
Accelerometer & ADXL377~\cite{adxl377} & \SI{540}{\micro\watt} \\ \hline
Microphone    & ADMP801~\cite{admp801} & \SI{17}{\micro\watt}   \\ \hline
Temperature   & LM94021~\cite{lm94021} & \SI{13.5}{\micro\watt} \\ \hline
\end{tabular}
\end{adjustbox}
\vspace{-2mm}
\end{table}

\subsection{Self-sustaining Sensor}
TunnelTags' main task is to sense physical phenomena. We couple the sensing and harvesting components to design what we call a self-sustaining sensor. It can track different physical phenomena, while operating by harvesting energy from ambient energy sources. Coupling the sensing and harvesting components has the advantage that some phenomena can also be a source of energy.  Thus, by enabling the sensor to both harvest energy and sense, we reduce the complexity and cost of its design. As an example, light can both be sensed and harvested using solar cells~\cite{bfvls}.

\fakepar{Sensors} TunnelTag supports diverse sensors to sense different phenomena such as ambient light conditions, temperature and vibrations. As the next step in the operation of the TunnelTag, i.e., processing sensor readings through the PPP, requires an analog signal, we restrict ourselves to sensors with analog outputs. To support operations on harvested energy, we look for sensors that sense at a low power consumption. We list such sensors in Table~\ref{sensorself}.

\fakepar{Light Sensor} We prototype and demonstrate a light sensing application as a concrete use case of our system~(Section~\ref{handgesture}). Consequently, we instantiate a light sensing self-sustaining sensor. This is, however, challenging, as some applications require sensing light at a high rate~\cite{li2016practical,li2017reconstructing}. As an example, some applications need to receive node identification information through visible light communication~(VLC), while others only passively track coarse changes in the light caused by shadow events such as hand gestures~\cite{solargest,bfvls}.
Hence, we design a light sensor that supports different sensing rates. We design the light sensor by fusing active and passive light receivers, which we describe next.

\input{sections/passivereceiver}

An active receiver consumes more energy than a passive receiver, but operates in diverse conditions (low light levels and high sensing rate). We design an active receiver using existing designs that involve coupling a photodiode with a transimpedance amplifier (TIA)~\cite{li2015human,li2016practical,li2017reconstructing}. The active receiver uses an SLD70-BG photodiode~\cite{photodiode-sld70-bg}, due to its responsiveness to the most commonly encountered light conditions. As TIA, we use a Texas Instruments OPA838~\cite{opa838} because of its large gain bandwidth product and low current consumption. 

A challenge with the active receiver is that 
its amplifier has a fixed gain
which limits the operating light conditions. On the other hand, visible light can be very dynamic due to natural changes or mobility.  To tackle this, we build on Wang et al.~\cite{jsaclight} and design a complimentary gain (high and low) active receiver. However, this requires us to devise a mechanism to select the appropriate receiver based on the light conditions. Hence,  we design the \emph{switchover} mechanism shown in  Figure~\ref{fig:lightdynamicsgain}.  The mechanism passively senses ambient light levels through the passive receiver, and selects the appropriate active receiver using a low-power comparator and multiplexer. Our switchover mechanism consumes a peak power of \SI{1}{\micro\watt} which represents a significant improvement over Wang et al.'s work~\cite{jsaclight} which required frequent ADC measurements on a microcontroller.

\fakepar{Harvesting Energy} To harvest energy we use the Texas Instruments BQ25570~\cite{bq25570} harvester that has also been used by others~\cite{powifi,bfvls,batteryfreecellphone}. We chose this harvester as it has a very low startup voltage and very low quiescent current.  To store energy we use the capacitor Seiko CPX3225A~\cite{capacitor-cpx3225a} of capacitance \SI{7.5}{\milli\farad}, one of the smallest super capacitors with sub-inch dimensions that is available on the market. The choice of the energy harvesting source depends on the phenomenon being tracked. For example, for light we use a solar cell for harvesting energy. For other sensors such as temperature we can use a thermoelectric generator and for vibrations a pizeoelectric element to harvest energy.

\input{sections/polymorphicdigitisation}

\begin{figure}[!t]
    \centering
    \includegraphics[width=\linewidth]{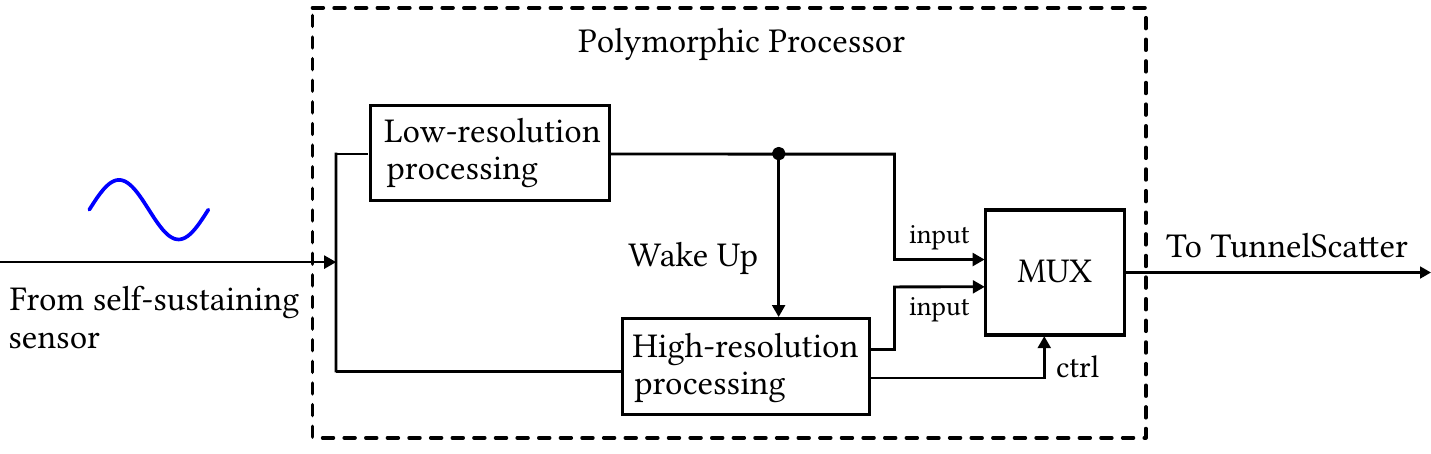} 
        \vspace{-4mm}
    \caption{\emph{Polymorphic Processing Pipeline }. It adapts its capabilities based on the  harvesting conditions, and processes using a low-resolution or a more capable but energy-expensive high-resolution component.  }
    \label{fig:polymorphicprocessor}
    \vspace{-4mm}
\end{figure}

\subsection{Low Power Transmissions using TunnelScatter}
\input{sections/ultralowpowercomm}

%% file: sections/passivereceiver.tex
A passive receiver is a sensor that can sense light conditions while consuming the least possible energy. We implement such a receiver using a solar cell building on our recent work~\cite{bfvls}. We select IXYS KXOB22~\cite{ixys-kxob22} and Powerfilm thinfilm solar cells~\cite{powerfilm} for sensing and energy harvesting. 
\begin{figure}[!t]
    \centering
    \includegraphics[width=0.8\linewidth]{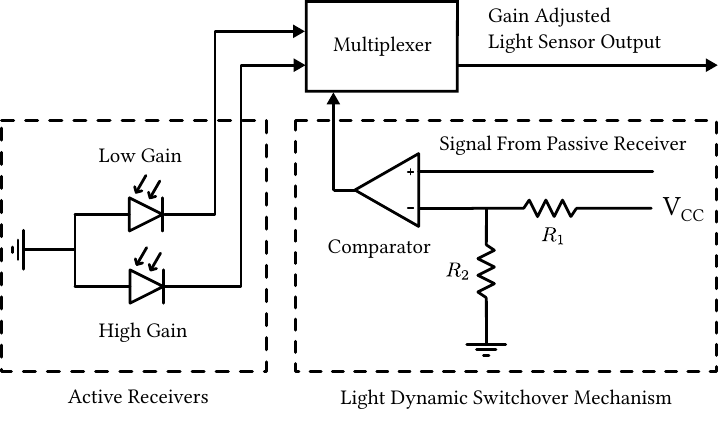} 
    \vspace{-4mm}
    \caption{\textit{Switchover mechanism.} We tackle the dynamics of changes in the light conditions by switching between receivers with different gain configurations.}
    \label{fig:lightdynamicsgain}
    \vspace{-6mm}
\end{figure}

%% file: sections/polymorphicdigitisation.tex
\subsection{Processing Sensor Readings}

\begin{figure}[!t]
\centering
\includegraphics[width=0.6\linewidth]{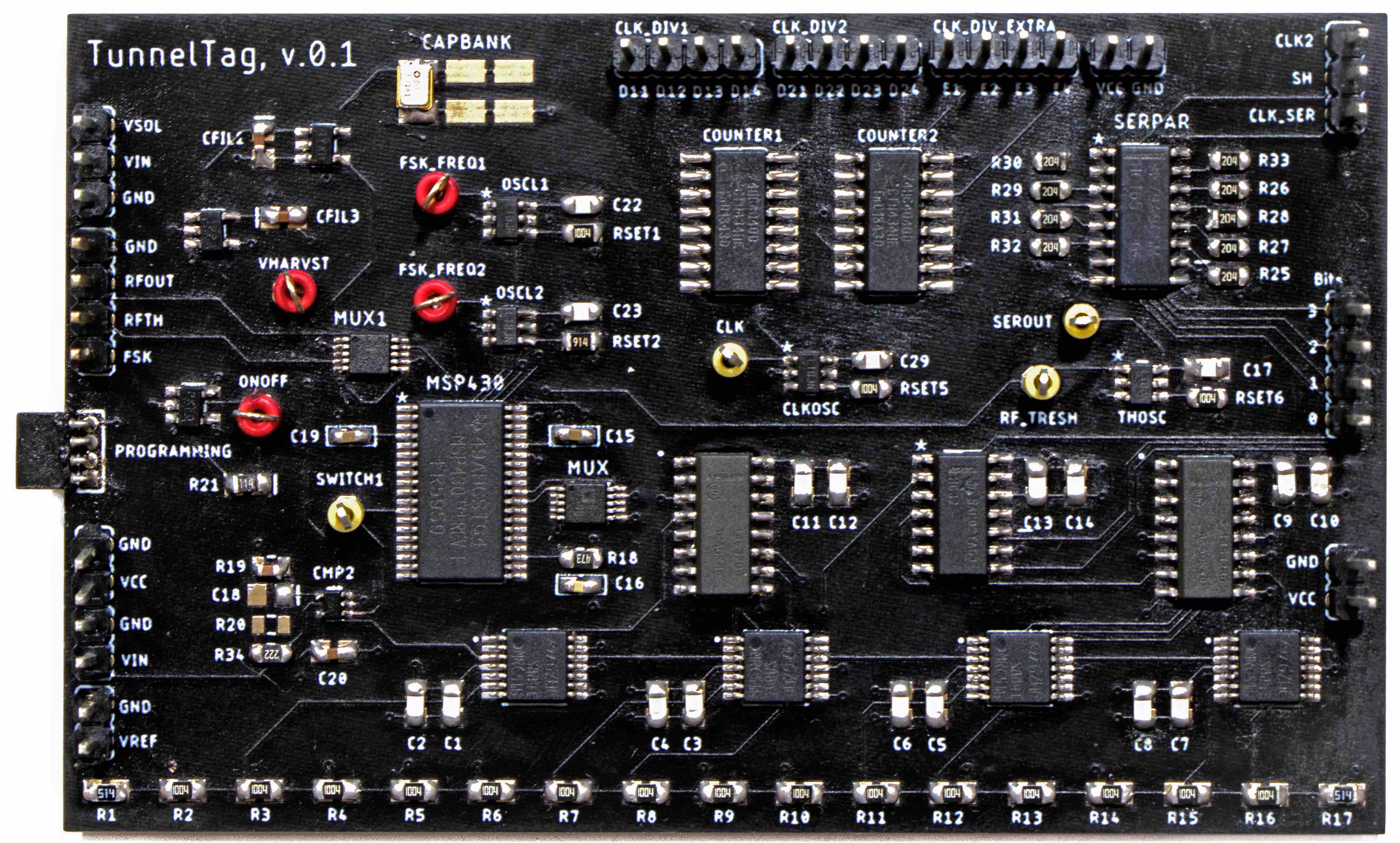}
    \vspace{-4mm}
\caption{\emph{Polymorphic Processing Pipeline  Prototype.}  The thresholding circuit acts as a wakeup mechanism for the energy-expensive microcontroller.} 
\label{fig:tunneltag}
    \vspace{-4mm}
\end{figure}

The acquired sensor readings need to be processed and digitised, before they are communicated to a powerful edge device. This step can be energy expensive, especially when operating on harvested energy. As a consequence, recent systems optimise this step by taking advantage of the very low power nature of backscatter communication.  These systems either optimise the processing block~\cite{ekhonet}, or backscatter without performing any processing like ADC operations~\cite{batteryfreecellphone,batteryfreehdcamera,bfvls}. We build on these systems, and design a component which we call polymorphic processing pipeline (PPP).  We show the hardware prototype in Figure~\ref{fig:tunneltag}.

PPP is based on the insight that in a typical deployment the energy harvesting conditions can significantly vary, as for example, light conditions change due to day and night. This impacts the processing operations that can be performed. Under poor harvesting conditions like low light levels, a lower sensing resolution and no processing might  conserve energy and ensure continuous sensing, and vice versa. PPP adapts its capabilities based on the prevailing energy harvesting conditions.  PPP consists of a low-resolution processing~(LRP), and a high-resolution processing~(HRP) component, as we show in Figure~\ref{fig:polymorphicprocessor}. The LRP acts like a wake-up mechanism for the comparatively energy-expensive HRP, a design inspired by wake-up radios~\cite{wakeup}.

\fakepar{Low-resolution Processing} When operating under challenging energy harvesting conditions, the PPP performs no processing, and digitises sensor readings at a low resolution in order to conserve energy. To support low-resolution digitisation, we build on the design of the thresholding circuits  used on backscatter tags to support reception~\cite{interscatter,wifibackscatter,ambientbackscatter} or RF sensing~\cite{allsee}. In its simplest form, a thresholding circuit consists of a low-power comparator coupled with a low-pass filter, which enables digitisation at sub-\SI{}{\micro\watt}. 

\begin{figure}[!t]
    \centering
    \includegraphics[width=0.9\linewidth]{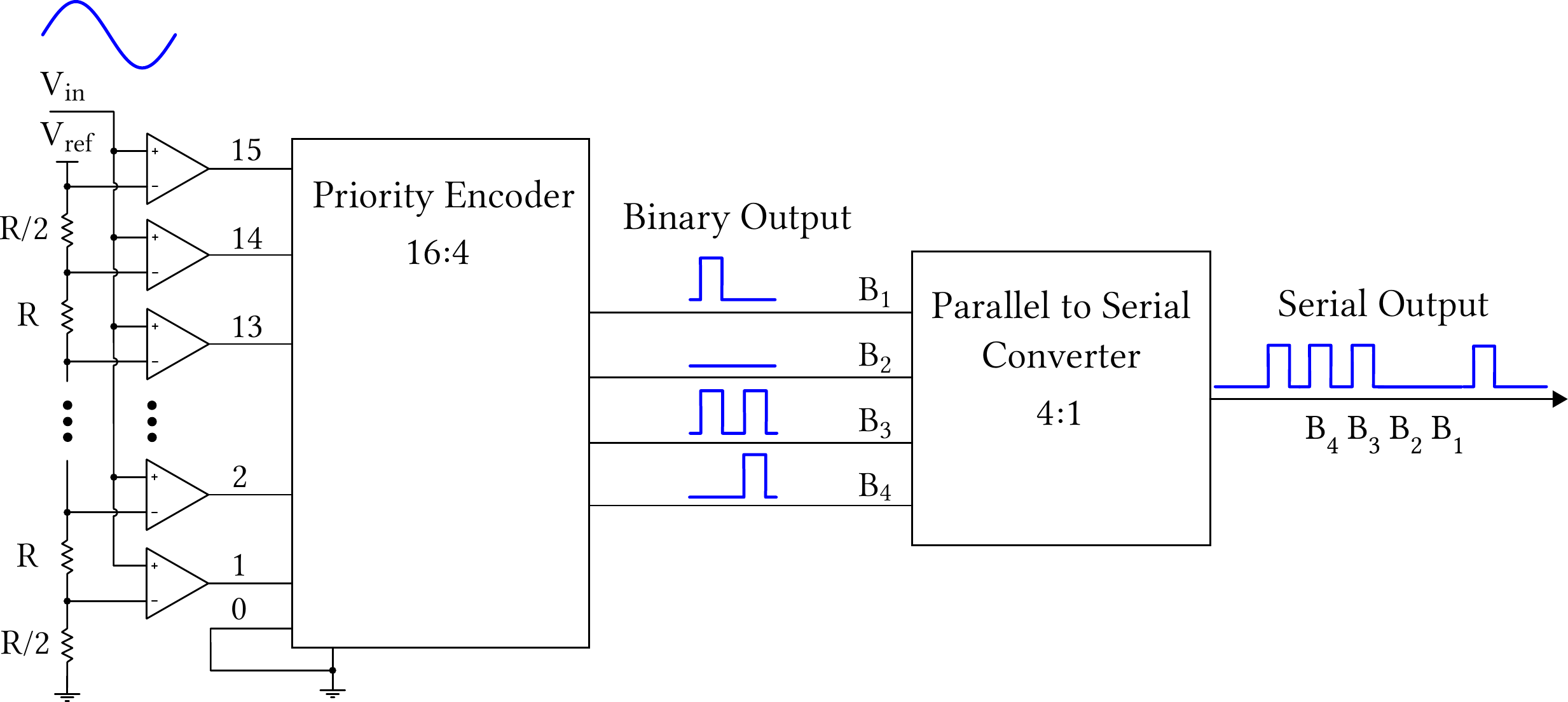} 
        \vspace{-4mm}

    \caption{\textit{Low-resolution Processing.} LRP is supported by a multibit thresholding  circuit. The circuit consumes 5.8-\SI{}{\micro\watt}s of power at a frequency of \SI{200}{\hertz}. }
    \label{fig:multibit}
    \vspace{-4mm}
\end{figure}

One limitation of thresholding circuits used in existing systems is the sensing resolution, as they act like a 1-bit ADC, which implies that some information is lost in the digitisation process. To overcome this limitation, we design a multibit thresholding circuit by building on the design of Flash ADC as shown in Figure~\ref{fig:multibit}. At a high level, the multibit thresholding circuit converts analog sensor readings to digitised bits through a series of comparators. We design the thresholding circuit to support 4-bit resolution. Due to the use of digital components,  comparators, and passive filters the mechanism has a very low power consumption.  The power consumption varies with sampling frequency, and consumes a  power of \SI{5.8}{\micro\watt}s at \SI{200}{\hertz}, a  frequency sufficient to support many sensing applications such as hand gesture recognition.

\fakepar{High-resolution Processing}  LRP trades off sensing resolution for power consumption. For applications that require a higher sensing resolution, and when energy harvesting conditions permit, the PPP supports digitisation at a higher sensing resolution using the low-power microcontroller~(MCU) MSP430~\cite{msp430} that has also been used by other energy harvesting systems~\cite{batteryfreecellphone,naderiparizi2015wispcam}. When there is an event of interest such as a hand gesture over the sensor, the LRP wakes up the MCU. Then the MCU senses the energy harvesting condition, and under sufficient conditions performs energy-expensive higher resolution digitisation~(up to 12 bit) and processing.

%% file: sections/ultralowpowercomm.tex
As the last step in the operation of TunnelTag, the digitised sensor readings are transmitted to an edge device for further processing using the TunnelScatter mechanism.

\begin{figure}[t]
\centering
\includegraphics[width=0.75\linewidth]{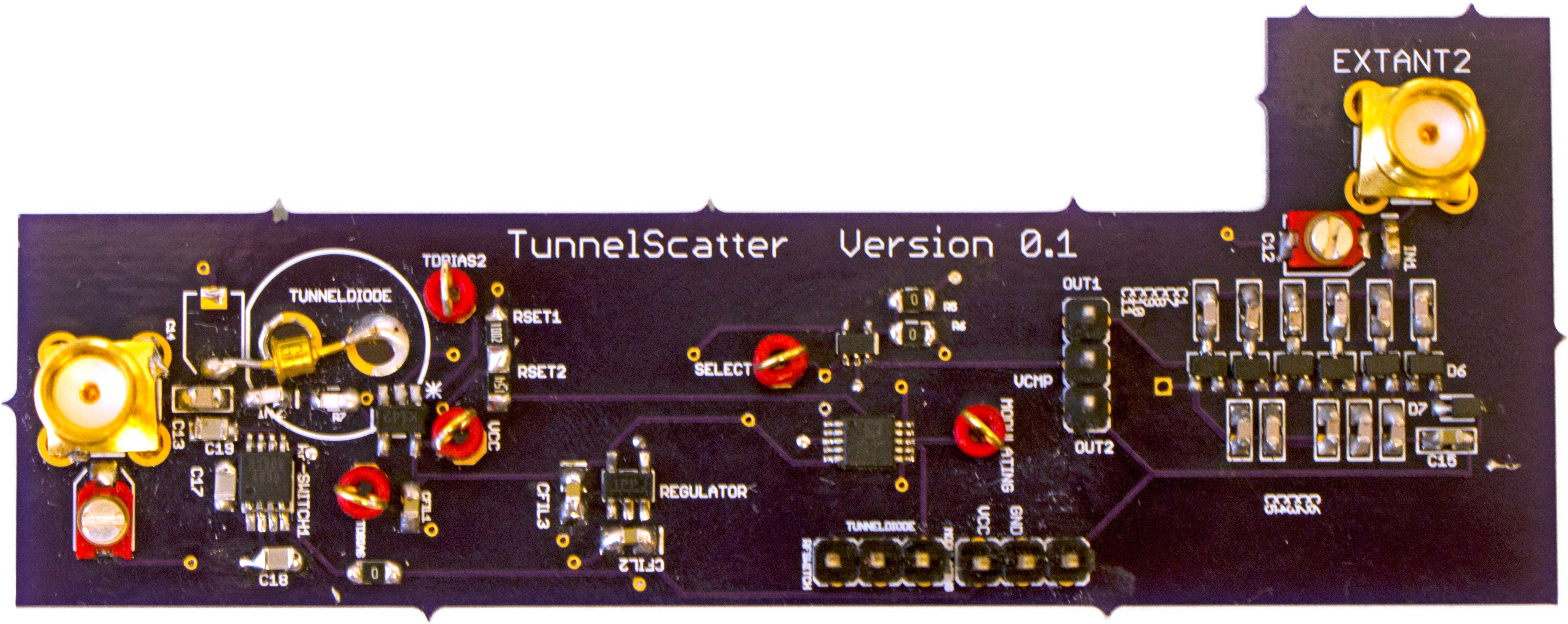}
\vspace{-4mm}
\caption{\emph{TunnelScatter  Prototype.} It enables communication in different ACS conditions, including when no ACS is present in the environment. }
\vspace{-4mm}
\label{fig:tunnelscatter}
\end{figure}

\fakepar{Overview} The TunnelScatter mechanism uses tunnel diodes to significantly enhance the design of existing backscatter tags~\cite{lorea,peng2018plora,hitchhike,passivewifi}. At a high level, the mechanism works as follows:  TunnelScatter passively senses and adapts to the strength of the ACS. In the presence of a strong ACS, TunnelScatter backscatters the signal using a conventional RF-switch similar to LoRea~\cite{lorea}. 
When the ACS is weaker, for example, when the tag is not close to the ACS emitter, the mechanism backscatters the weak ACS with a notable gain using a tunnel diode as a reflection amplifier~\cite{amato2017achieving}, thereby achieving a significant improvement in communication range. Finally, in the absence of an ACS, the mechanism generates a signal locally using tunnel diodes, modulates the signal with sensor readings, and transmits it. Thus, TunnelScatter enables communication in diverse ACS conditions. We show the TunnelScatter hardware prototype in Figure~\ref{fig:tunnelscatter}. Before discussing the detailed design of Tunnelscatter, we provide a brief background on tunnel diodes.

\begin{figure}[t]
\centering
\includegraphics[width=0.9\linewidth]{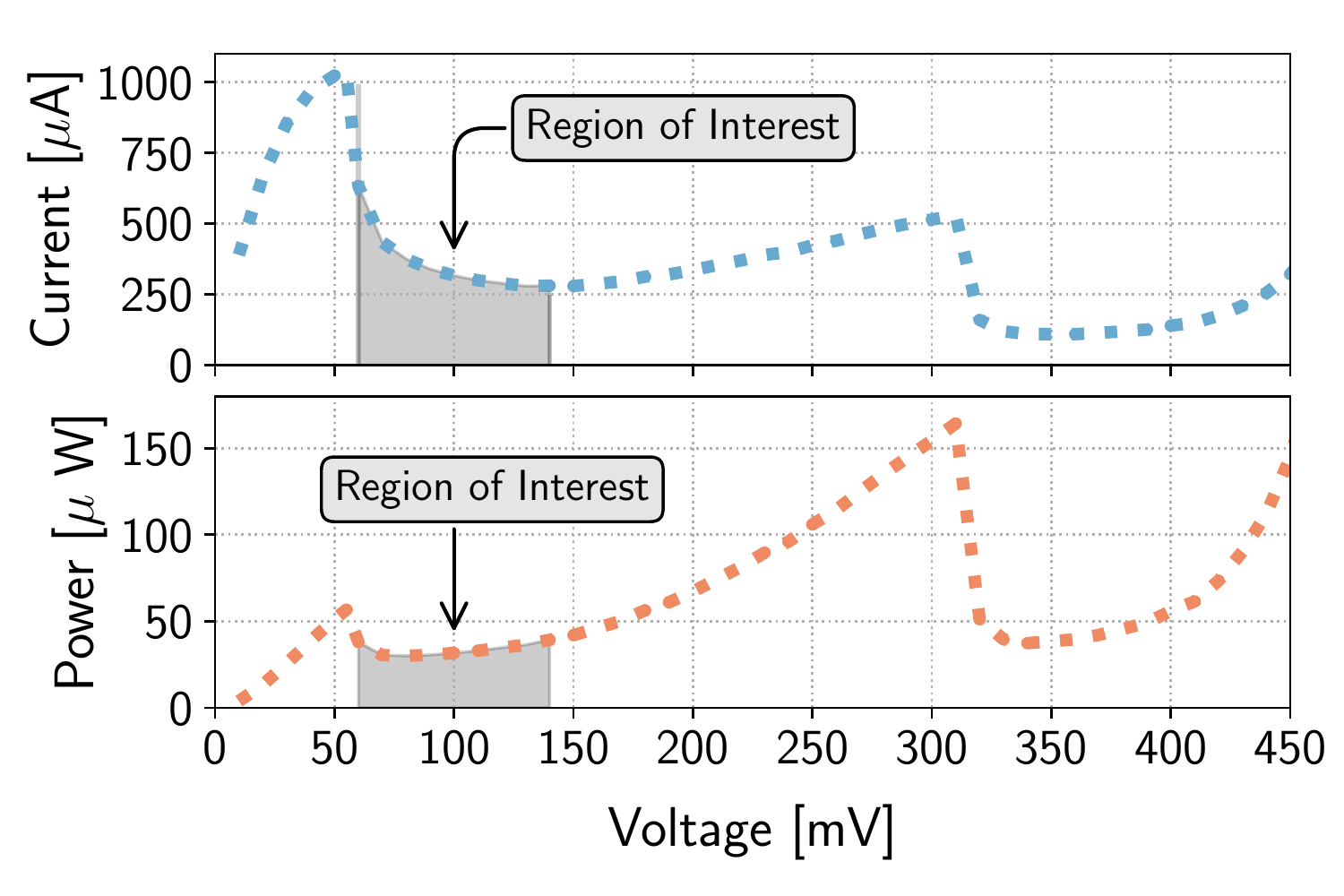}
\vspace{-6mm}
\caption{\textit{Tunnel diode characterization}. We bias and operate the tunnel diode within a subset of the region of negative resistance, which we call region of interest.}
\vspace{-4mm}
\label{fig:tdiv}
\end{figure}

\fakepar{Tunnel Diodes} 
A tunnel diode is a two terminal device with a p-n junction that has an order of magnitude higher doping concentration than the junctions of conventional diodes. Therefore, tunnel diodes have a very thin depletion region at the junction. Due to the thin depletion region, tunnel diodes demonstrate an effect called quantum tunneling effect~\cite{rca1963rca}. This effect causes a region of negative resistance, i.e., as we increase the voltage beyond the peak voltage, the current through the device decreases, as we show in Figure~\ref{fig:tdiv}. 
The region of negative resistance 
makes it possible to design tunnel diode-based RF amplifiers and oscillators,  as we demonstrate in this paper.  In this paper, we use a tunnel diode GE 1N3712~\cite{1n3712}, due to its low peak voltage~(\SI{0.65}{\milli\volt}), and current consumption~(\SI{1}{\milli\ampere}). We bias the tunnel diode to a voltage between \SI{65}{\milli\volt} and \SI{150}{\milli\volt} to keep it within the region of interest (RNR). We derive the negative resistance~($R_{RNR}$) from the slope of the IV-curve within the highlighted region of interest in Figure~\ref{fig:tdiv}.
\begin{equation}
    R_{RNR} = \frac{1}{\text{slope}_{RNR}} 
\end{equation}
For our case, $R_{RNR}$ computes to approximately \SI{-287}{\ohm}. 

\subsubsection{Ambient Carrier-less Transmissions}\hspace*{\fill} 

As discussed earlier, a major obstacle that hinders the  deployment of backscatter systems is the requirement of a strong ACS. Conventional  transceivers do not encounter this problem, as they generate the carrier signal locally. However, these transceivers are too energy expensive for battery-free sensor tags since they consume~\SI{}{\milli\watt}s of peak power~\cite{passivewifi,ensworth2015every} due to the use of active analog components such as RF oscillators.  Hence, to transmit without an ACS, TunnelScatter requires a low-power oscillator operating at  hundreds of \SI{}{\mega\hertz} while consuming \SI{}{\micro\watt}s of power.  Operation at hundreds of \SI{}{\mega\hertz} ensures that we can communicate in the commonly used ISM bands.  On the other hand, as discussed earlier state-of-the-art backscatter systems design or use \SI{}{\micro\watt}s oscillators operating at tens of \SI{}{\mega\hertz}~\cite{freqshift}. TunnelScatter tackles this challenge through a tunnel diode-based RF oscillator. 

\begin{figure}[t!]
    \centering
    \includegraphics[width=0.45\textwidth]{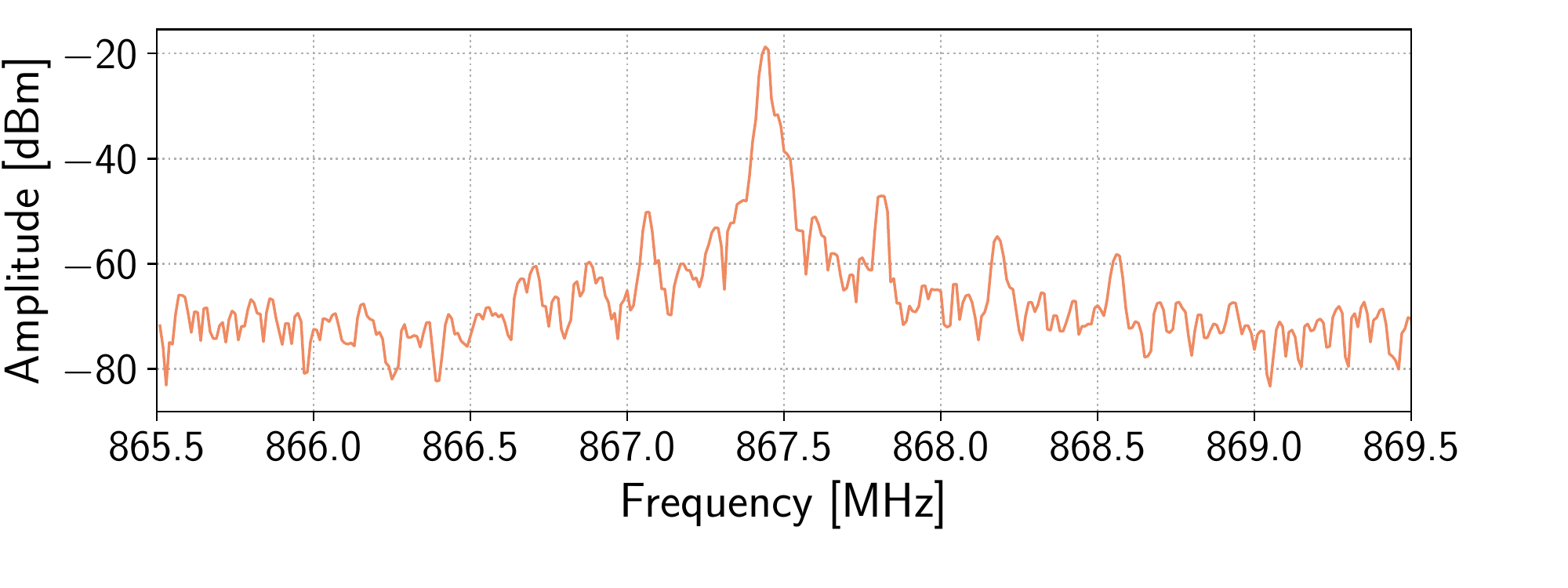}
    \vspace{-4mm}
    \caption{\emph{Spectrum of TDO}. We generate a signal in the 868 MHz band, at a peak biasing power of \SI{57}{\micro\watt}. }
    \label{fig:tdplot}
    \vspace{-4mm}
\end{figure}

\fakepar{Tunnel diode oscillator} In ACLT mode, TunnelScatter locally generates a signal using a tunnel diode oscillator. The signal is then modulated using amplitude shift keying~(ASK). We design the TDO to operate at a frequency band of 868~\SI{}{\mega\hertz}, a license-free band for communication in major parts of the world. The TDO can also be tuned to operate at other bands. We design the TDO taking advantage of the fact that RNR enables tunnel diodes to oscillate at high RF frequencies~\cite{rca1963rca,gemanual}. In fact, tunnel diodes were used to design RF oscillators more than half a century back~\cite{rca1963rca}.  However, they are not widely used due to their limited peak current which restricts the output power~\cite{gemanual}. We use the limited power consumption to our advantage to design RF oscillators for battery-free sensor tags.

\begin{figure*}[t!]
\centering
\includegraphics[width=0.9\linewidth]{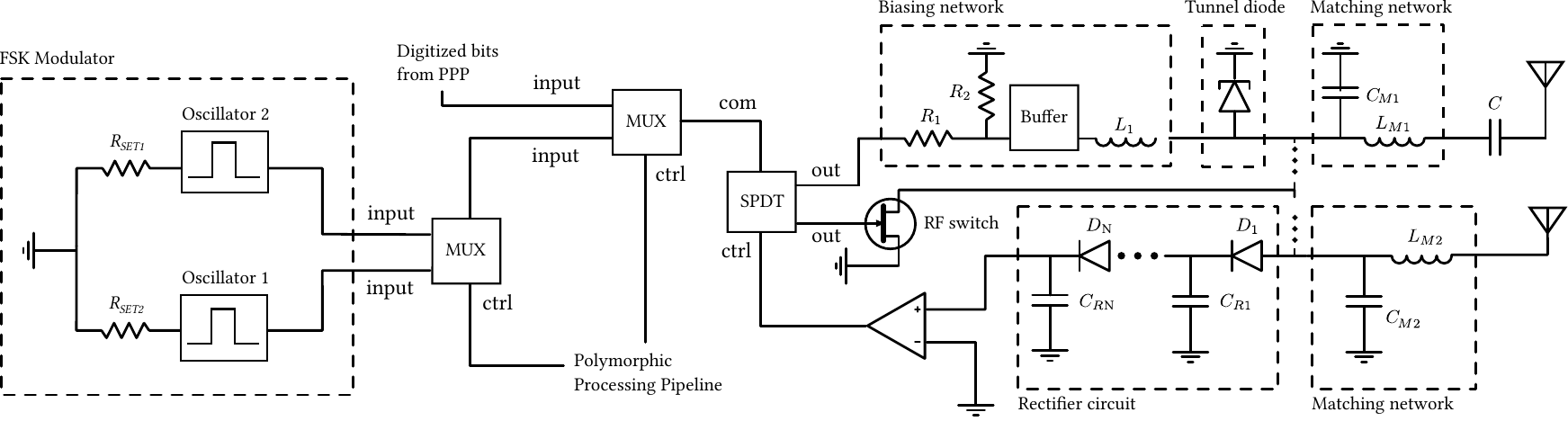}
\vspace{-6mm}
\caption{\emph{TunnelScatter schematic.} TunnelScatter uses the TDO to operate in diverse ACS strengths (including in the absence of an ACS). The TunnelScatter mechanism employs a passive envelope detector to select between the tunnel diode and the RF-switch by sensing the strength of the ACS. This selection mechanism ensures a high SNR of received backscatter transmissions at the edge device.}

\vspace{-5mm}
\label{fig:quantumscatter}
\end{figure*}

\begin{figure}[t!]
    \centering
    \includegraphics[width=\linewidth]{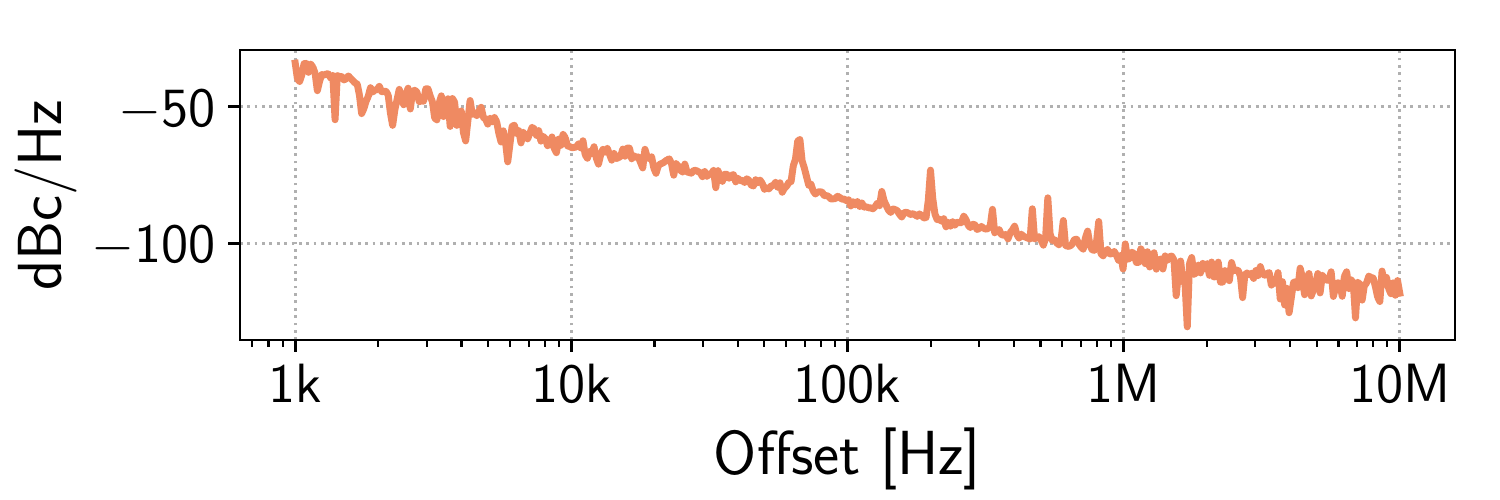}
    \vspace{-6mm}
    \caption{\emph{Phase noise}. We trade off a higher phase noise for low power consumption with a tunnel diode oscillator when compared to precision oscillators.}
    \label{fig:phasenoise}
    \vspace{-6mm}
\end{figure}

We show the schematic of the TDO in the top right part of  Figure~\ref{fig:quantumscatter}. The TDO is designed using a tunnel diode, a biasing circuit, a matching network, and an antenna. The matching network~($C_{M1}$ and $L_{M1}$) sets the resonant oscillating frequency. The biasing circuit configures the tunnel diode into the RNR which is essential to enable oscillations.

 \fakepar{TDO performance} We evaluate the TDO, as it dictates the communication ability of the TunnelScatter mechanism. Like any wireless system, the TDO can be affected by ambient noise or interfering signals, and hence we perform the measurements in an anechoic chamber.  First, we connect the TDO to a Keithley 2810 RF signal analyzer~\cite{keithely} through a cable, and capture the spectrum plot. Figure~\ref{fig:tdplot} shows the result of the experiment. It shows that most of the energy is contained within the resonant frequency of the TDO, i.e, \SI{867.4}{\mega\hertz}. We note that the resonant frequency itself can change with bias voltage or can drift slightly over time, as we evaluate in Section~\ref{eval:act}. The peak strength of the signal generated by TDO was \SI{-19}{\decibel}m. 
 
 Phase noise is commonly used to characterize the performance of oscillators. We investigate the phase noise using the RF signal analyzer. In Figure~\ref{fig:phasenoise}, we observe that we have a higher phase noise as compared to precision oscillators, which is expected as we trade off 
 some stability for power consumption. As we demonstrate in Section~\ref{evaluation} this does not impact reception negatively.

\subsubsection{Amplified Backscatter Transmissions}\hspace*{\fill} 
\label{sect:tdra}

In many scenarios, an ACS might be present. 
State-of-the-art backscatter systems~\cite{lorea,lorabackscatter,peng2018plora,passivewifi,hitchhike,freerider,freqshift} achieve a large range only when the ACS is sufficiently strong. TunnelScatter overcomes this limitation  through the  amplified backscatter transmitter~(ABT) mode. The ABT mode takes advantage of the fact that in the presence of an ACS, the TDO demonstrates a behaviour called called injection locking~\cite{amato2017achieving,injectionlocking,rca1963rca}. This allows the TDO to backscatter the ACS, and achieve a significant gain also with a weak ACS.

\fakepar{Injection locking} This is a phenomenon where an oscillator is influenced by a signal from another oscillator operating in the vicinity of the resonant frequency of the first oscillator. Injection locking as a phenomenon was discovered in the late 16th century by Christiaan Huygens, who observed two pendulum clocks on the wall synchronizing with each other. This phenomenon is used in many transceivers today~\cite{histinjection}.

\begin{figure*}[t]
\centering
\includegraphics[width=0.8\linewidth]{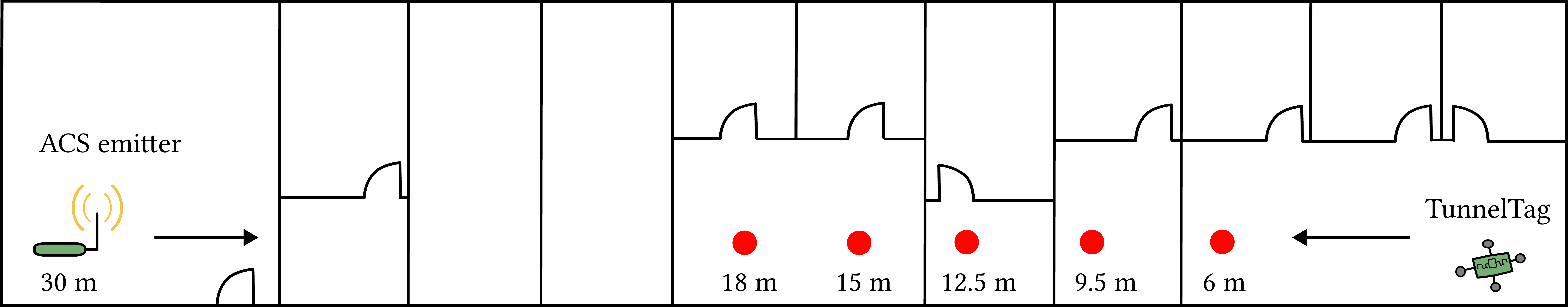}
\vspace{-3mm}
\caption{\emph{Layout for the indoor experiments.} Red dots indicate the location of the receiver. The ACS emitter was switched off for the experiments involving TunnelScatter in ACLT mode. The distance is from the TunnelTag.}
\vspace{-6mm}
\label{indoorlayout}
\end{figure*}

\begin{figure}[t!]
    \includegraphics[width=0.8\linewidth]{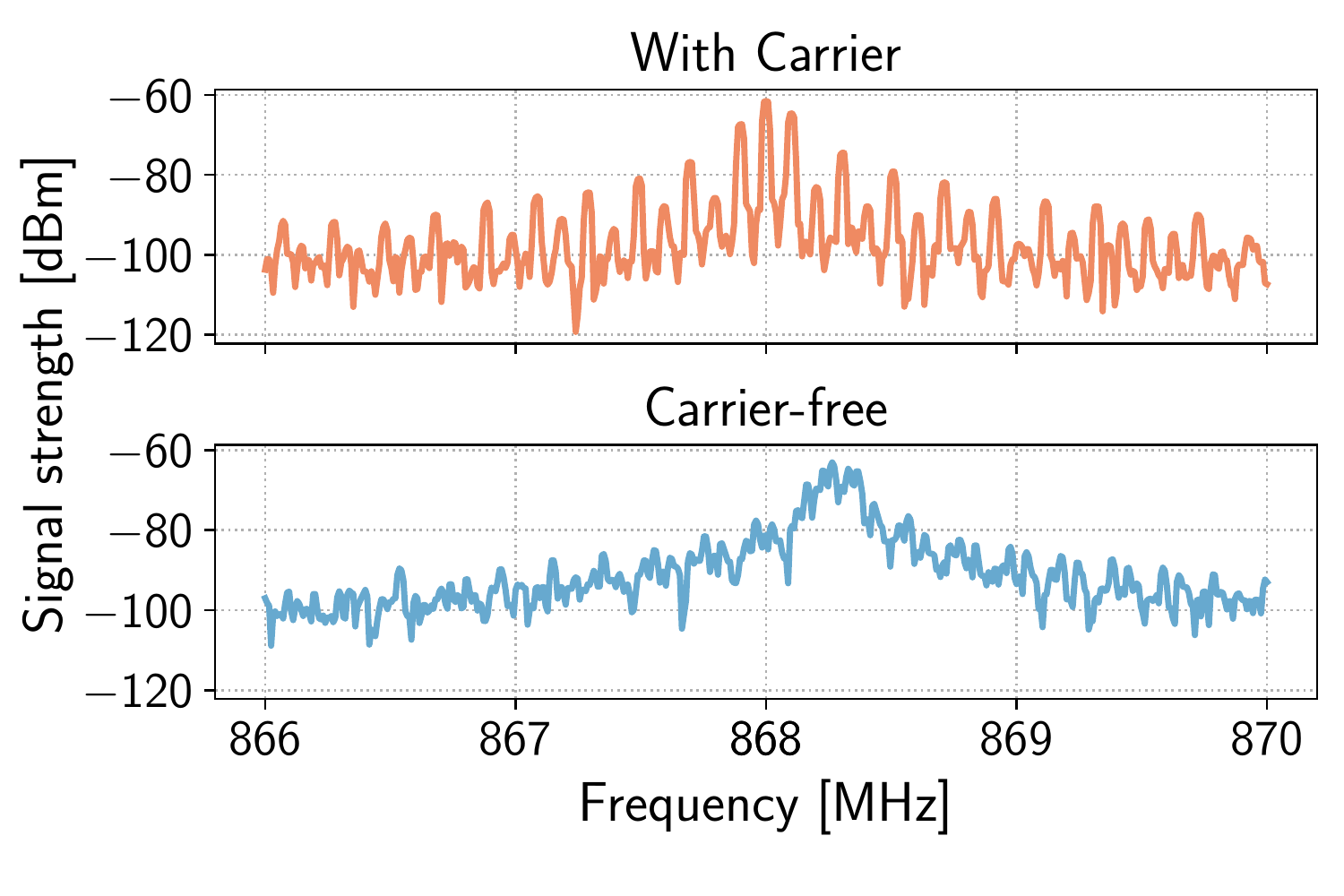}
    \vspace{-6mm}
    \caption{\emph{Injection locking spectrum}. TDO due to injection locking process latches onto an ACS.}
    \vspace{-6mm}
    \label{fig:latching}
\end{figure}

We use the TDO's injection locking to backscatter a weak ACS with significant gain which makes the tunnel diode act as a reflection amplifier~\cite{amato2018tunneling,amato2018tunnel2,amato2015long,amato2017achieving}. In this mode, the TDO uses some energy to bias the diode to amplify, modulate and reflect back the ACS.  To demonstrate this phenomenon, we set up TunnelTag approximately \SI{1}{\meter} away from an ACS emitter~(SDR), and co-locate an RF spectrum analyzer. First, we observe the signal generated from the TDO in absence of an ACS. Next, we generate a carrier signal using the SDR. Figure~\ref{fig:latching} shows that the TDO latches onto the impinging signal, and is influenced by the ACS. In fact, this also enhances the harmonics produced by 
the backscatter process.  Injection locking also reduces the phase noise of the oscillator~\cite{injectionlocking}. In our experiments, we have also observed that the injection locking ability of the TDO is influenced by the strength of the ACS, and the offset of the ACS from the resonant frequency of the TDO.  We observe that the stronger the external signal is, the further it can be from the resonant frequency of TDO, and yet result in injection locking. Due to space constraints, we do not present this result.

\fakepar{Amplified Backscatter Transmitter} We design the ABT mode by building on the injection locking of TDO and the FSK modulator presented by LoRea~\cite{lorea}. Since the tunnel diode acts as a reflection amplifier, this mode can achieve a significant range improvement over LoRea when reflecting a weak ACS. We show our high level circuit design in Figure~\ref{fig:quantumscatter}.  It works as follows: we generate the two frequencies to enable FSK transmissions, using low-power oscillators, LTC 6906~\cite{ltc6906}. One of these signals is selected using a multiplexer according to the bit being transmitted. Next, this signal is fed to the biasing network which tunes the tunnel diode into the negative resistance region. In the presence of an ACS, it gets modulated and reflected back with a gain. We have measured the gain using a vector network analyzer~(VNA), and found it to be as high as \SI{35}{\decibel} at \SI{-60}{\decibel}m of input power.

\fakepar{Switchover mechanism} While the ABT achieves a significant improvement over conventional tags~\cite{lorea}, we observe on a VNA that the gain of the tunnel diode reflection amplifier is impacted by the incident carrier signal strength. This has also been observed by  Amato et al.~\cite{amato2018tunneling}.  We believe, one reason for this behaviour is the change in the impedance of the tunnel diode. The impedance of the tunnel diode is a function~\cite{amato2018tunneling,chang2025microwave} of frequency~(f), the bias voltage~($V$), RF input power~($P_{RF}$), and temperature~(T), i.e.,  
$Z_L(f, V, P_{RF}, T)$. This means, as we increase the strength of the incident carrier signal, $P_{RF}$ increases, which causes the impedance of the tunnel diode to change, which we, and Amato et al.~\cite{amato2018tunneling} 
have also observed.  On the other hand, the matching network~($C_{M1}$ and $L_{M1}$) of the circuit is configured for a specific impedance value of the tunnel diode.  Hence, there is a mismatch which contributes to the poor performance of the tunnel diode at higher ACS strength, which makes a conventional tag perform better when the ACS is stronger.

We overcome the above challenge with a  switchover mechanism that we show in Figure~\ref{fig:quantumscatter}. We build this mechanism using a passive envelope detector that is also used on backscatter tags~\cite{ambientbackscatter,livingiot}.  We use a passive envelope detector to sense the ACS' strength, and based on the ACS strength, we switch between tunnel diode or the standard RF-switch~(HMC190BMS8~\cite{rf-switch}). In designing this mechanism, we take advantage of a limitation of envelope detectors, which is their poor sensitivity. Envelope detectors are designed using discrete components, and commonly have a sensitivity of approximately \SI{-40}{\decibel}m~\cite{livingiot,braidio}.  The poor sensitivity of envelope detectors ensures that they only output a signal when the ACS signal is sufficiently strong, which activates the conventional backscatter tag. On the other hand when the ACS is weak, the mechanism selects the tunnel diode. We design the switchover mechanism using an ultra-low power multiplexer and comparator~(TS 881~\cite{ts881}).

%% file: sections/eval.tex
In this section, we evaluate different aspects of our system in a range of  conditions. The key highlights are:

\begin{itemize}
\item TunnelScatter enables communication through several walls in a non-line-of-sight environment in ACLT mode, using the tunnel diode oscillator.
\item TunnelScatter allows communication through several floors of a building while backscattering~(ABT mode) a weak ACS. In a similar setting, a tag similar to LoRea~\cite{lorea} achieves a range of only \SI{3}{\m}.
\end{itemize}

\subsection{Ambient Carrier-less Transmitter}
\label{eval:act}
In this section, we evaluate the ability of \systemnosc to communicate using the ACLT mode, i.e., in the absence of an ACS. 
Our experiments focus on the stability of the  signal generated by the TDO, and the  range achieved in a challenging non-line-of-sight~(NLOS) scenario.

\fakepar{Setup} To measure the stability of the TDO, we connect the TunnelScatter to an RF spectrum analyzer; Keithley 2810~\cite{keithely}. We have seen that the TDO can be influenced by ambient RF noise, and hence we perform these measurements in an anechoic chamber. The ACLT mode transmits amplitude-modulated transmissions. To receive the transmissions, a CC1310 radio acts as energy detector (RSS sampling), similar to FS-backscatter~\cite{freqshift}. We perform the experiment in our office, as shown in Figure~\ref{indoorlayout}. The walls between the rooms consist of insulated gypsum and are approximately 16 cm thick. The rooms are equipped with standard office equipment such as chairs, tables with drawers, and monitors. 

\begin{figure}[t!]
    \includegraphics[width=\linewidth]{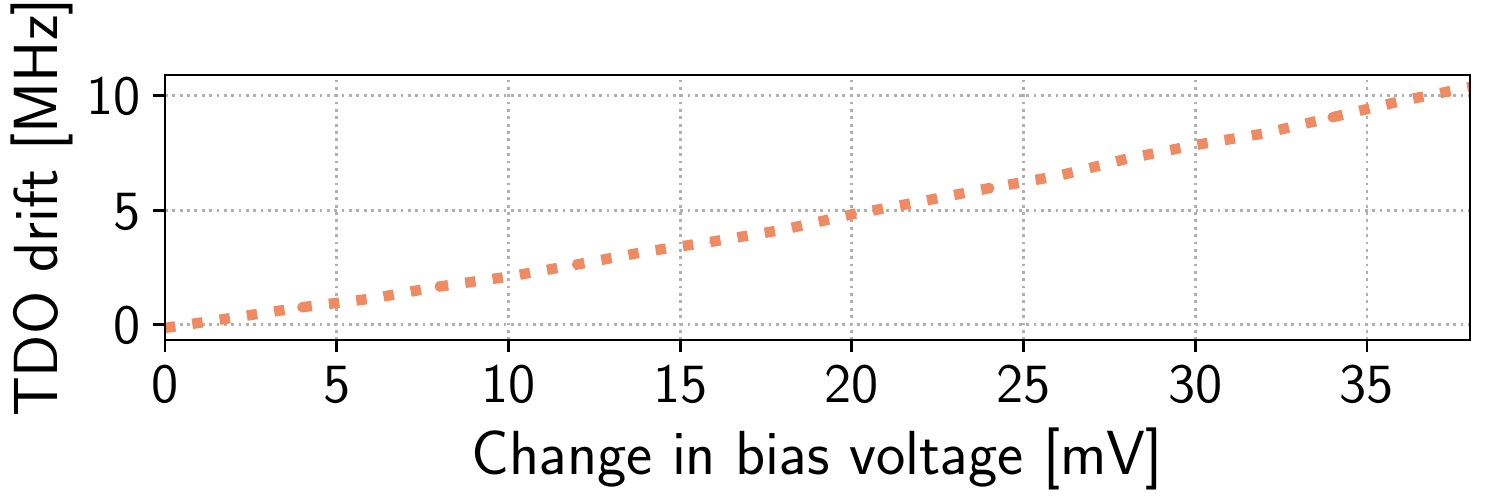}
    \vspace{-4mm}
    \caption{\emph{Drift with voltage}. The resonant frequency of the TDO changes linearly with the bias voltage.}
    \vspace{-6mm}
    \label{fig:tdvoltage}
\end{figure}

\fakepar{Oscillator stability with bias voltage} To enable oscillations, the tunnel diode has to be biased to the negative resistance region, as shown in Figure~\ref{fig:tdiv}. However, within this region, we observe that the frequency of the TDO is influenced by the bias voltage.  To investigate this closer, we connect the biasing network of the TunnelScatter mechanism to an external waveform generator~(Digilent Analog Discovery 2) which enables us to control the bias voltage. We change the bias voltage, and observe the frequency of the TDO. We plot the results as the change in the bias voltage, and corresponding changes observed in the frequency of the TDO in Figure~\ref{fig:tdvoltage}. Our results show that the frequency of the TDO changes linearly with the bias voltage. To counter this drift, we maintain a constant bias voltage on TunnelTag using a ultra-low power regulator~(S-1313~\cite{s1313}).

\begin{figure}
    \includegraphics[width=\linewidth]{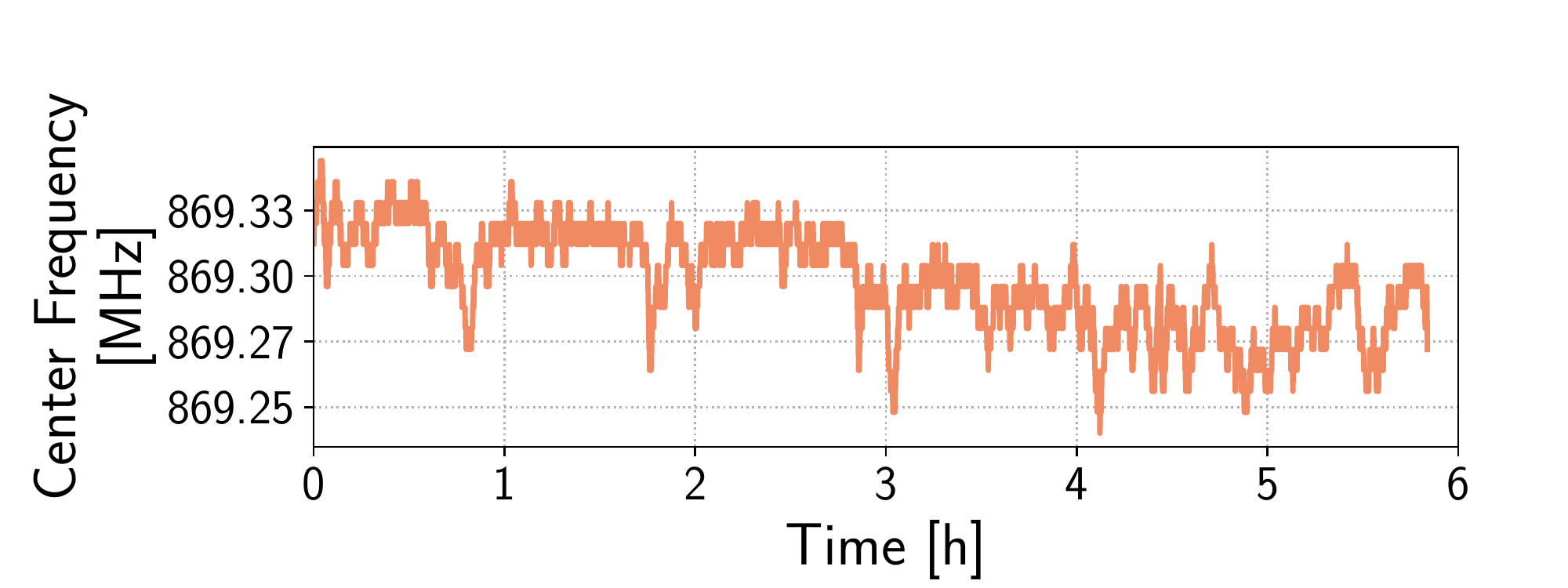}
    \vspace{-6mm}
    \caption{\emph{TDO drift}. The frequency of the TDO drifts slightly~(tens of \SI{}{\kilo\hertz}) over the period of six hours.}
    \vspace{-6mm}
    \label{fig:tdosc-drift}
\end{figure}

\fakepar{Oscillator stability over time} In this experiment, we look at the long term stability of the TDO. As we consume significantly lower power compared to commercially available precision oscillators, we expect the TDO to be less stable. We provide a constant bias voltage to the TDO using a low-power regulator. We keep track of the frequency of the TDO at an interval of \SI{6}{\second}. We run the experiment for a duration of \SI{6}{\hour}. Figure~\ref{fig:tdosc-drift} demonstrates the result of the experiment. Throughout the experiment, the frequency of the TDO varies slightly, but remains within \SI{80}{\kilo\hertz} of the initial frequency, with a standard deviation of \SI{19}{\kilo\hertz}.  This is not a problem for our system, as we receive transmissions using ASK and energy detectors, which are less impacted by small shifts in the frequency of the carrier signal. Further, in our experiments, we have noticed that under injection locking in ABT mode, the backscatter signal remains stable with little deviation, which enables reliable reception.

\begin{figure}[t!]
    \centering
    \includegraphics[width=\linewidth]{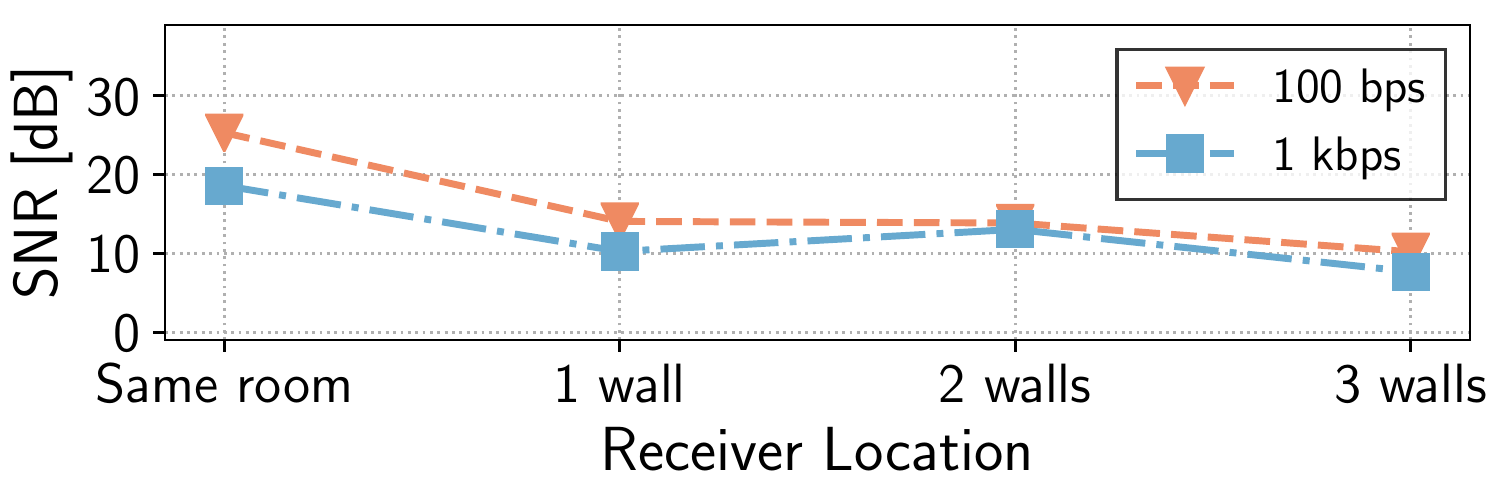}
    \vspace{-4mm}
    \caption{\emph{ACLT through the wall}. We can communicate through several walls covering a distance of \SI{18}{\meter} when using ACLT mode of TunnelScatter.}
    \label{fig:snr}
    \vspace{-6mm}
\end{figure}
 
\fakepar{Through the wall communication} In this experiment, we investigate the ability of \systemnosc to support communication without an ACS.  Due to space constraints, we present only a range experiment. The receiver is placed at different distances from the tag.  Our results in Figure~\ref{fig:snr} show that we are able to communicate through three walls despite the NLOS environment, at bitrates of 100 bps and 1 kbps. We cover a distance of \SI{18}{\meter}. We observe slightly anomalous behaviour around the second wall, which might be caused by the presence of metallic equipment in this location. As we had seen earlier, the TDO's output power is restricted to approx.~\SI{-19}{\decibel}m. Yet, we are able to  communicate through the wall due to the high sensitivity of the CC1310 receiver.

\subsection{Amplified Backscatter Transmitter}
\label{commability}
We evaluate the ability of TunnelScatter to support backscatter transmissions. A line-of-sight~(LoS) environment  significantly improves the range of backscatter systems~\cite{lorea} and our system is no exception. Hence, we  focus on challenging NLoS environments. 
We perform experiments indoors within our university building.

\fakepar{Setup} We use an SDR, USRP B200, as an ACS emitter. We equip both the SDR and the TunnelTag with a VERT900 antenna~\cite{vert900}. For receiving the transmissions, we use the FSK mode on the CC1310, with parameters similar to those used by LoRea~\cite{lorea}. We use a frequency shift of \SI{100}{\kilo\hertz} to avoid self-interference~\cite{lorea}, and bitrate of 2.9 kbps. We compare the ABT results with a tag similar to LoRea. We have verified that this tag achieves a communication range similar to the one reported for LoRea.

\fakepar{Multifloor communication} We evaluate the range improvement of TunnelScatter when backscattering with amplification with an ACS incident on the tag of a very low  strength.  We place the tag at a distance of \SI{1}{\meter} from the ACS emitter~(SDR) but configure the SDR to generate an ACS of strength of \SI{-27}{\decibel}m, which represents orders of magnitude lower strength compared to existing long range systems~\cite{lorea,lorabackscatter,peng2018plora}. First, we measure range and reliability with a LoRea tag. Then we replace it with a TunnelTag.

\begin{figure}[!t]
    \centering
    \includegraphics[width=\linewidth]{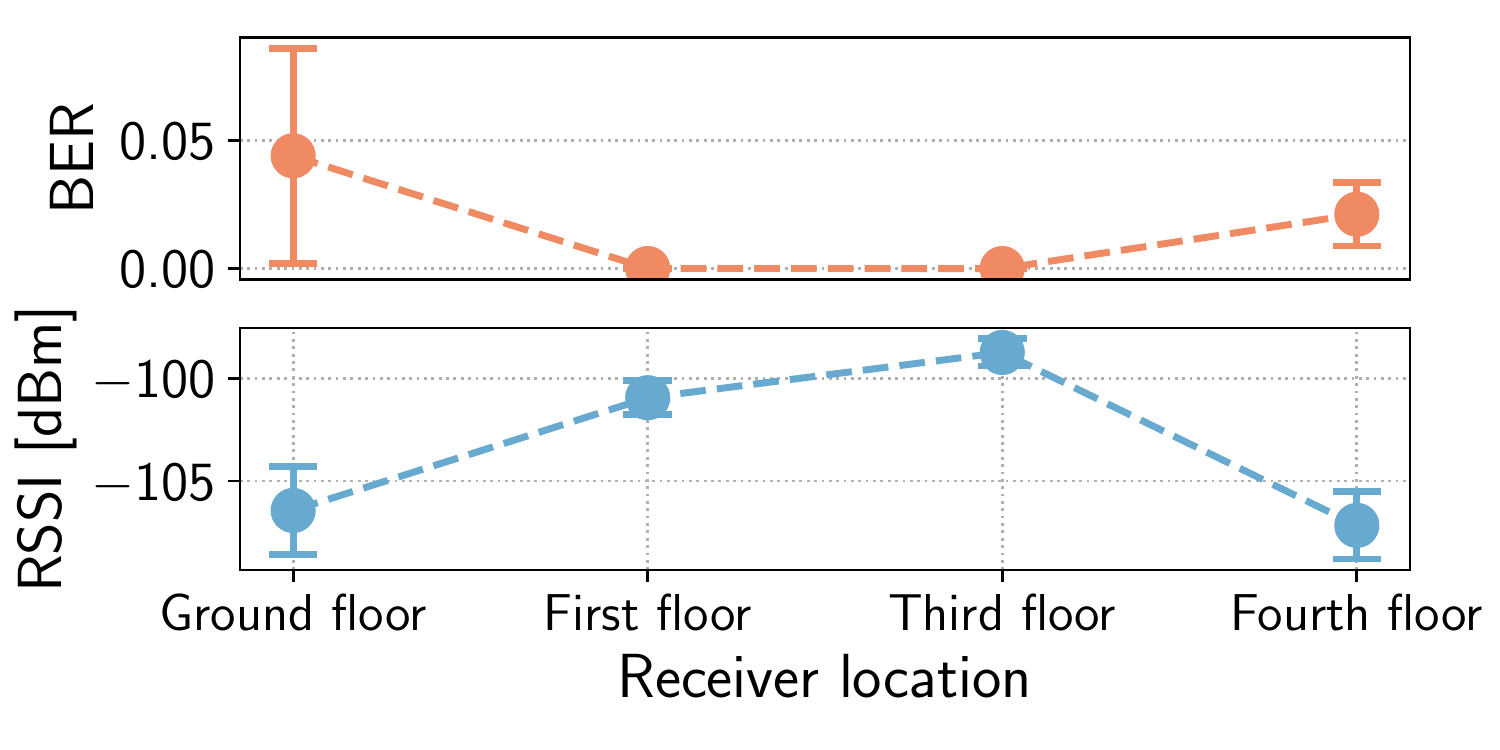} 
    \vspace{-6mm}
    \caption{\emph{Multifloor backscatter.} With ABT, we communicate to all four floors of our university building when backscattering a weak ACS. }
    \label{multifloor_tunnel_diode}
    \vspace{-6mm}
\end{figure}

Figure~\ref{multifloor_tunnel_diode} demonstrates the result of the experiment. With the TunnelTag, we can communicate easily through four floors of the university building, while under similar settings, the LoRea tag achieves a  range of only three meters. The experiment demonstrates that the TunnelScatter mechanism achieves a significant range improvement over the state-of-the-art backscatter LoRea tag~\cite{lorea}.

\begin{figure}[!t]
    \centering
    \includegraphics[width=\linewidth]{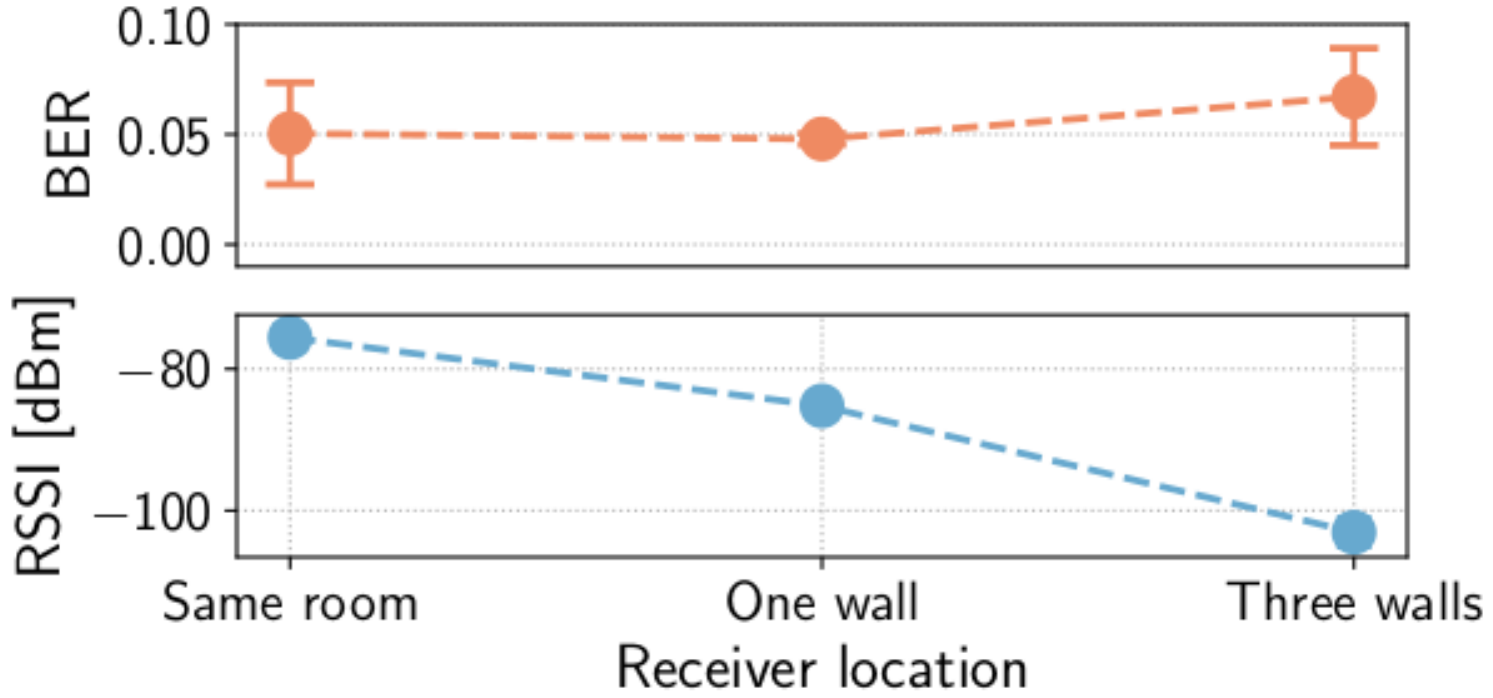} 
    \vspace{-6mm}
    \caption{\emph{Through the wall backscatter.} Even when the ACS emitter is placed several walls away~(\SI{30}{\meter}), TunnelScatter enables the tag to communicate \SI{15}{\meter}, going through several walls. }
    \label{cgkitchen_tunnel_diode}
    \vspace{-4mm}
\end{figure}

\fakepar{Through the wall communication} Next, we evaluate the ability of TunnelScatter to operate in a challenging NLOS environment. We set up a carrier generator transmitting with a signal of strength \SI{16}{\decibel}m, located \SI{30}{\meter} away, separated by seven walls from the tag as shown in Figure~\ref{indoorlayout}. We experiment similar to the multifloor experiment described above. Figure~\ref{cgkitchen_tunnel_diode} demonstrates the result. The LoRea tag is able to communicate only a few meters, and cannot communicate through the wall while the TunnelTag communicates through three walls covering a distance of \SI{15}{\meter}.

\fakepar{Switchover mechanism} We evaluate the advantage of switching between a tunnel diode and a conventional RF switch for backscatter transmissions. This switchover happens depending on the strength of the ambient carrier signal incident on the tag. In the experiment, we place the TunnelTag at a distance of \SI{1}{\meter} from the ACS emitter. We co-locate the tag with a spectrum analyzer to keep track of the incident ACS strength. We position a CC1310 receiver about \SI{4}{\meter} from the tag. We keep track of the signal strength of the received transmissions at the CC1310 receiver.

\begin{figure}[!t]
    \centering
    \includegraphics[width=\linewidth]{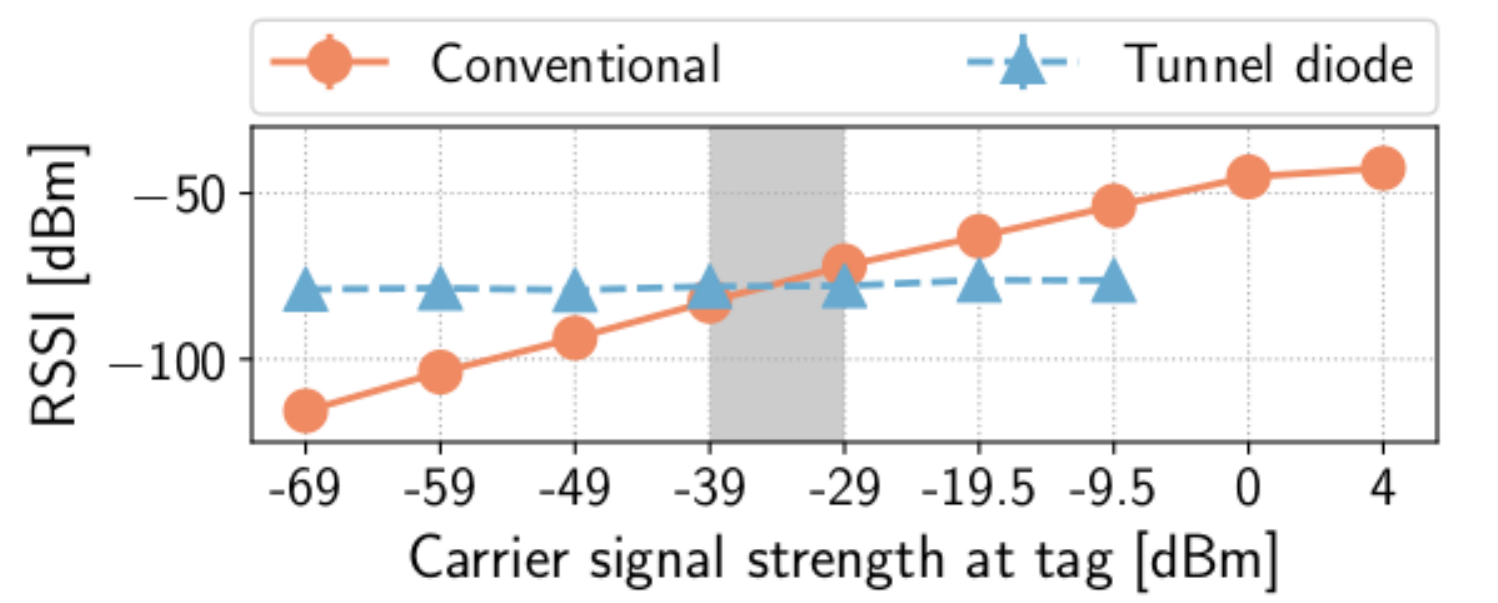} 
    \vspace{-6mm}
    \caption{\textit{Tunnel diode switchover mechanism}. As we increase the strength of the ACS, the conventional tag outperforms the tunnel diode tag. Shaded regions represent the sensitivity levels of the envelope detector.}
    \label{fig:tdraswitchover}
    \vspace{-4mm}
\end{figure}

Figure~\ref{fig:tdraswitchover} demonstrates the results of the experiment. The figure depicts that the strength of the backscattered signal from the conventional tag increases with the strength of the ACS, similar to other backscatter systems~\cite{lorea}. The figure  shows that the strength of the backscattered signal remains constant with the tunnel diode-based tag. At an ACS strength between \SI{-30}{\decibel}m and \SI{-40}{\decibel}m, the conventional tag starts to outperform the tunnel diode-based tag. 
 
As described in Section~\ref{sect:tdra}, using the input from the passive envelope detector, \systemnosc switches between tunnel diode and the standard RF switch. This makes certain that \systemnosc adapts to the strength of the ambient carrier signal, and ensures that the SNR of the backscattered signal is maximized at the receiver.

\subsection{Power Consumption}

We evaluate the power consumption of our system. As the ACS emitter and the edge device would be  powered externally, we do not evaluate their power consumption.

\fakepar{Experiment setup} To measure the power consumption, we operate using a \SI{2}{\volt} supply, as it decreases with the operating voltage~\cite{freqshift}. We measure the power consumption by connecting a  Fluke multimeter in series to the circuit to measure the current draw. To measure the power consumption of the TunnelScatter mechanism, we use a highly sensitive Keysight E36300 voltage supply~\cite{keysight}.

\fakepar{Polymorphic Processing Pipeline} The thresholding circuit's power consumption is dependent on the sampling frequency. At a  frequency of \SI{1}{\kilo\hertz}, the thresholding circuit comsumes \SI{11.4}{\micro\watt} of power. On the other hand, to support higher sensing resolution, PPP adapts under favorable energy harvesting conditions and uses the MSP430 microcontroller that consumes more power, as we show in Table~\ref{tab:adcpower}. We note that the power consumption is low enough to  support many sensing events, such as in hand gesture sensing.

\begin{table}[!tb]
    \centering
    \caption{\emph{PPP Power Consumption.} The thresholding circuit ensures that the energy expensive higher resolution processor is activated only during an event. }
    \vspace{-4mm}
    \begin{adjustbox}{width=\columnwidth,center}    \begin{tabular}{|l|l|l|}
        \hline 
         \parbox[t]{3cm}{Sampling frequency}  & \parbox[t]{3cm}{4-bit thresholding\\power consumption} & \parbox[t]{3cm}{MSP430 \\ power consumption} \\ \hline
         \SI{200}{\Hz} & \SI{5.8}{\micro\watt} & \SI{385}{\micro\watt} \\ \hline
         \SI{500}{\Hz} & \SI{7.9}{\micro\watt} & \SI{559}{\micro\watt} \\ \hline
         \SI{1}{\kHz} & \SI{11.4}{\micro\watt} & \SI{687}{\micro\watt} \\ \hline
    \end{tabular}
    \end{adjustbox}
    \label{tab:adcpower}
    \vspace{-6mm}
\end{table}

\fakepar{TunnelScatter} Due to the to low biasing voltage required by the tunnel diode, we measure the power consumption through the highly sensitive Keysight voltage supply~\cite{keysight}. We observe that the tunnel diode consumes a peak biasing power of \SI{57}{\micro\watt}, as we show in the Figure~\ref{fig:tdiv}. This low power consumption for biasing is similar to other works that have used a  tunnel diode as reflection amplifier operating at \SI{5}{\giga\hertz} band~\cite{amato2017achieving,amato2018tunneling,amato2018tunnel2,amato2015long}. The very low power consumption facilitates battery-free operation, while significantly improving the ability to operate at a distance far away from the ACS emitter.  The comparator, and standard RF-backscatter switch,  each consume sub-\SI{}{\micro\watt} of power. The envelope detector consumes no additional power. On the contrary, it can be used to harvest energy.

\fakepar{Self-sustaining light sensor} This sensor consists of the solar cell, active receivers~(high and low gain), and the  mechanism to select the receivers. The power consumption is influenced by the operation of the sensor which is application-dependent. At the lowest power state, the self-sustaining sensor consumes no energy, and only senses using the passive receiver which does not involve any amplification stage. This mode enables application scenarios such as the hand gesture recognition. To support high sensing rate or low light conditions, the self-sustaining sensor uses the active receiver. Depending on the light conditions, only one of the active receivers is enabled using the switchover mechanism. The active receiver has a peak power consumption of \SI{2.4}{\milli\watt}. Finally, the switchover mechanism consumes  \SI{1}{\micro\watt}. 

\subsection{Self-sustaining Sensor}

\fakepar{Harvesting energy} A feature of the self-sustaining sensor is its ability to operate battery-free, through harvested energy. 
The choice of energy source for harvesting depends on the application scenario. In this paper, we focus on the use of ambient light and solar cell for harvesting.  To evaluate the ability to harvest energy from indoor light, we connect the output of the energy harvester to a logic analyzer and vary the  light conditions. We keep track of the time taken to charge the capacitor to the maximum voltage supported by the harvester, and the operation time of the active light receiver operational under this condition, as this represents the most energy-expensive component of our system.

\begin{figure}[t]
    \centering
    \includegraphics[width=0.9\linewidth]{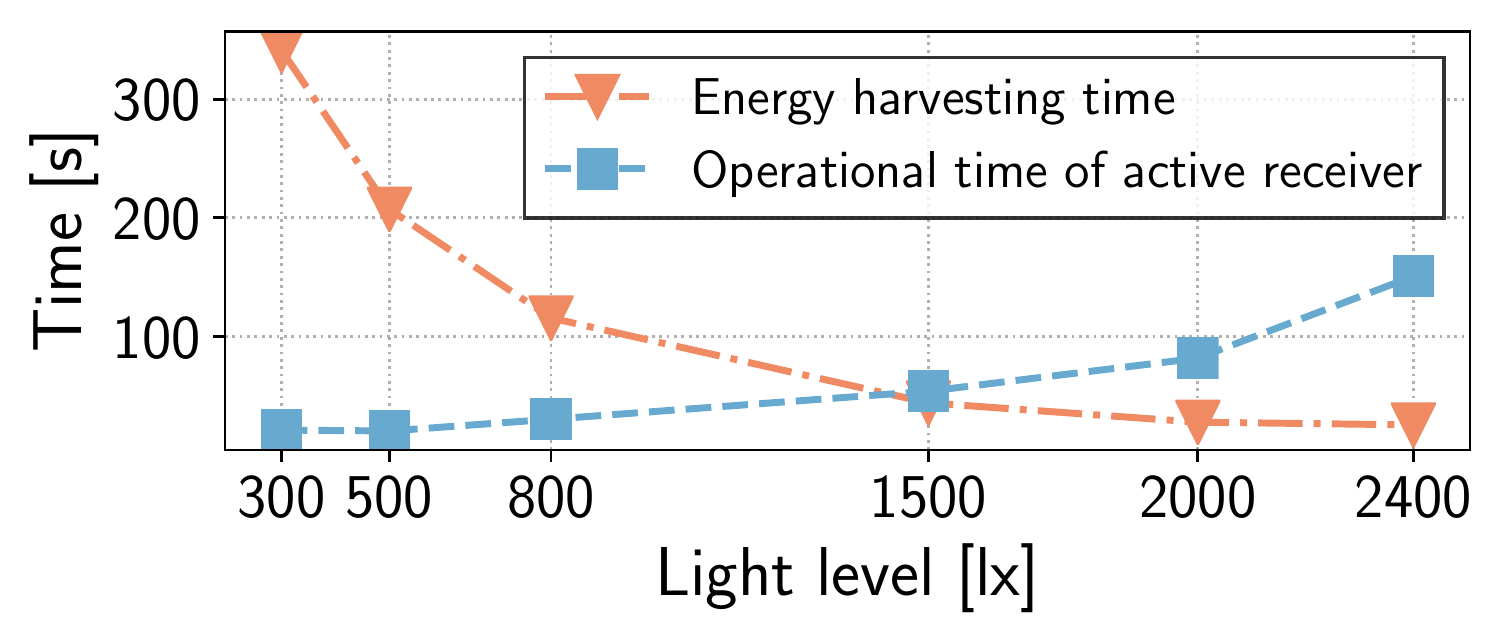}
    \vspace{-4mm}
    \caption{\emph{Self-sustaining light sensor performance.} Within a few minutes, we harvest enough energy under the most commonly found light conditions to operate the active light receiver for tens of seconds.}
    \label{fig:harvestingexp}
    \vspace{-6mm}
\end{figure}

Figure~\ref{fig:harvestingexp} shows the result of the experiment. We observe that we can charge the capacitor within a few minutes under light conditions that can be expected in an indoor environment. As expected, the harvesting rate increases with the light levels. Further, we can keep the active receiver operational on the harvested energy for tens of seconds. We note that while it takes a few minutes to charge the capacitor, once charged, the capacitor maintains its charge, and only uses the energy when an event such as a hand gesture is detected. Handling of the events is triggered by the energy-inexpensive passive receiver. 

\begin{figure}[t!]
    \centering
    \includegraphics[width=0.9\linewidth]{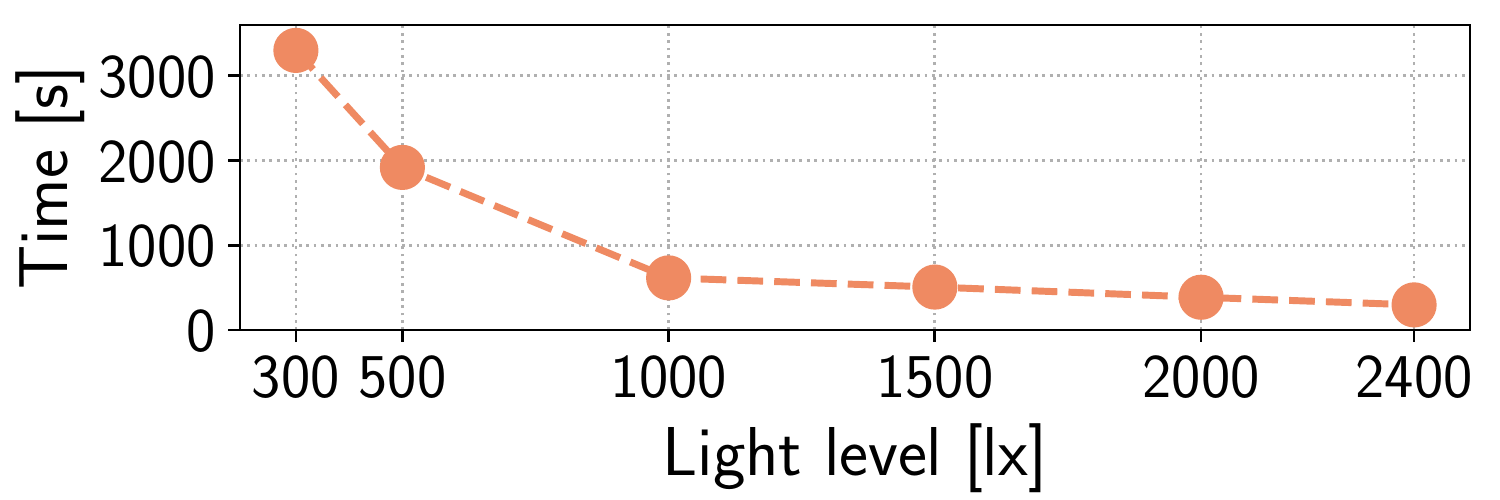}
    \vspace{-2mm}
    \caption{\emph{Cold start mode}. Cold start takes a significant time in low light conditions, but usually has to be performed once in an application deployment. }
    \label{fig:coldstart}
    \vspace{-5mm}
\end{figure}

\fakepar{Cold start mode} Until the harvester reaches a critical voltage~(\SI{1.8}{\volt}), it operates in a less efficient cold start mode. During this period, it takes a significant time to harvest energy. Once the cold start critical voltage is reached, the harvester charges at the faster rate shown in Figure~\ref{fig:harvestingexp}.  The operation in cold start mode needs to be performed only occasionally. We measure the time taken for the harvester to reach cold start voltage. Figure~\ref{fig:coldstart} shows that, as expected, the time to reach the cold start is quite high for low light levels but decreases drastically under 
bright light conditions. To decrease the time spent in cold start mode, we can increase the size of the solar cell or employ a small external power source.

\section{Application Use Case}
\label{appusecase}
We prototype and demonstrate visible light sensing-based hand gesture recognition as a concrete application enabled by TunnelScatter.
Moreover, we briefly discuss other sensing applications that can be enabled using TunnelScatter. 

\subsection{Hand Gesture Sensing}
\label{handgesture}

We investigate the feasibility to implement a hand gesture sensing system using TunnelScatter. We build on the vision of LiSense~\cite{li2015human,li2016practical} and Aili~\cite{li2017reconstructing}. We  imagine a scenario where one could control various devices in a home by sensing changes in light conditions caused by hand gestures, and communicate these changes through TunnelScatter.  In this section, we focus on the ability  to sense the changes in the light levels caused by hand gestures.  

\fakepar{Sensitivity}  We define the sensitivity of the light sensor as the minimum light level that produces sufficient voltage from the light sensor to enable digitization. This voltage is dictated by the thresholding circuit and in our architecture is above \SI{20}{\milli\volt}. To measure sensitivity, we vary the light levels and track them using a Texas Instrument sensor-tag~\cite{sensortag}. Next, using a logic analyser, we look for the particular light levels that cause the output of the light sensor to fall below the critical voltage of \SI{20}{\milli\volt}. As the sensitivity levels depend on the gain value used in the amplifier, we perform the experiment for the high~(\SI{900}{\kilo\ohm}) and low gain~(\SI{100}{\kilo\ohm}) settings.

Table~\ref{lightsensitivity} demonstrates the result of the experiment. We observe that the high-gain configuration can detect low light levels, but results in the amplifier being saturated easily. On the other hand, the low-gain configuration of the active receiver fails to detect low light levels, but gets saturated first at a light level of \SI{2200}{\lux}, which is very bright for indoor environments. The passive receiver detects the signal above the light level of \SI{30}{\lux}. Hence, it is important to fuse these individual receivers to tackle light dynamics. 

\begin{figure}[t!]
    \centering
    \includegraphics[width=0.8\linewidth]{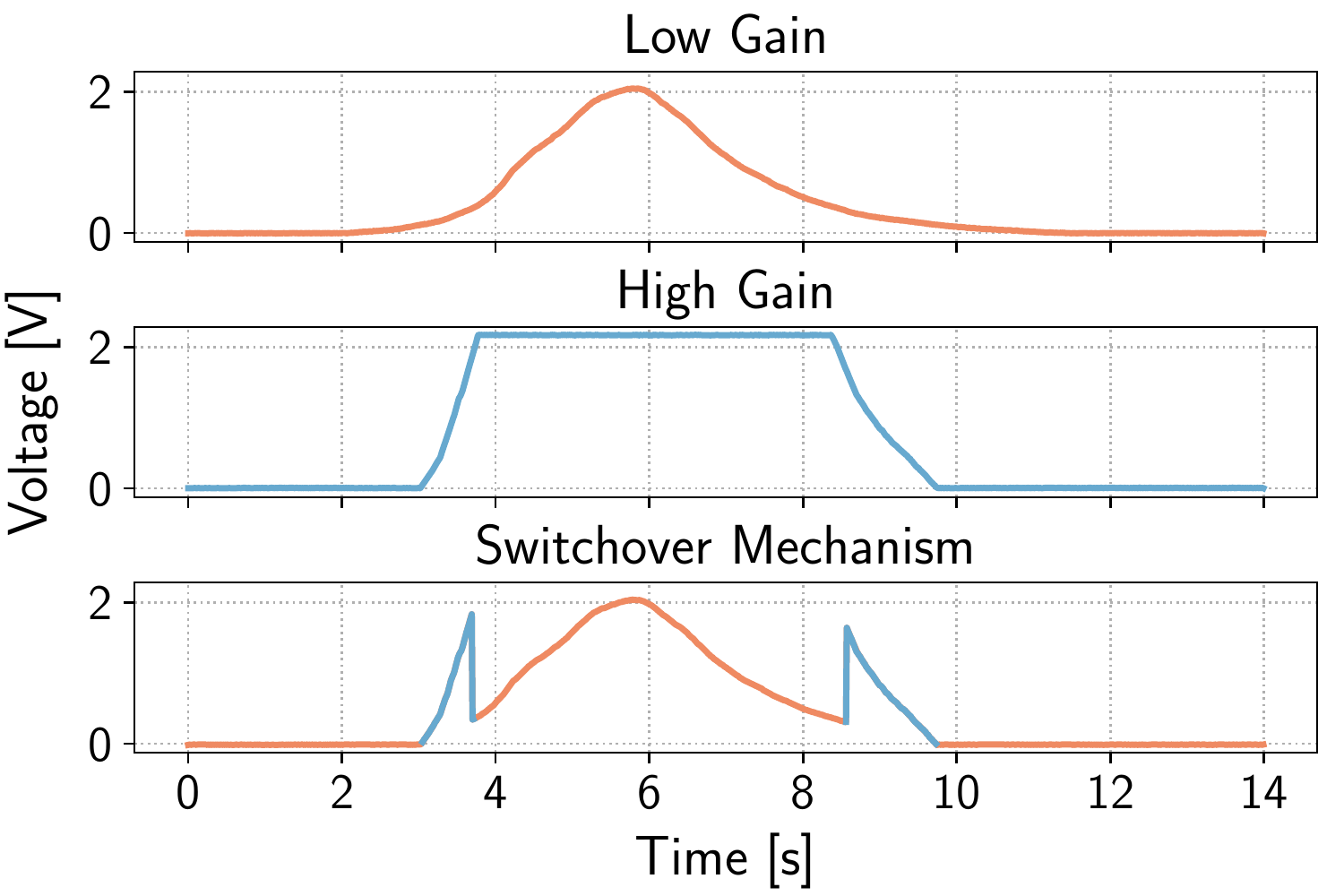}
    \vspace{-4mm}
    \caption{\emph{Handling light dynamics}. As the high-gain receiver saturates,  the switchover mechanism automatically switches to the low-gain receiver ensuring we can operate across diverse light conditions.}
    \label{fig:autogain}
    \vspace{-4mm}
\end{figure}

\begin{table}[!tb]
\centering
\caption{Operating light conditions for the light sensor in high and low gain configuration of active receiver.}
\label{lightsensitivity}
    \begin{adjustbox}{width=\columnwidth,center}  
\begin{tabular}{|l|l|l|}
\hline
\multicolumn{1}{|c|}{\begin{tabular}[c]{@{}c@{}}Configuration (k$\Omega$)\end{tabular}} & \multicolumn{1}{c|}{\begin{tabular}[c]{@{}c@{}}Minimum light (lx)\end{tabular}} & \multicolumn{1}{c|}{\begin{tabular}[c]{@{}c@{}}Maximum light  (lx)\end{tabular}} \\ \hline
High gain (900)                                                                                     & \textless 30                                                                                        & 350                                                                                                 \\ \hline
Low gain (100)                                                                                      & 100                                                                                                 & 2200                                                                                                \\ \hline
\end{tabular}
\end{adjustbox}
\vspace{-5mm}
\end{table}

\fakepar{Tackling dynamics} The previous experiment has shown that the receivers have different operating conditions. 
We use this observation to tackle the dynamics of the light conditions. The switchover mechanism autonomously switches to the appropriate receiver depending on the light conditions to maximize the SNR of the signal  from the light sensor.

To test the switchover mechanism, we perform the experiment under varying light conditions.  We change the ambient light conditions between  \SI{30}{\lux} to \SI{2500}{\lux}. We keep track of the output of the -- low gain, high gain -- active receiver, and also the output from the switchover mechanism.  Figure~\ref{fig:autogain} demonstrates that at  low light levels, the high-gain receiver gives a signal with high SNR, and is chosen by the switchover mechanism. However, as the light levels increase, the high-gain receiver saturates, and the switchover mechanism selects the low-gain active receiver. Finally, as the light levels decrease, the switchover mechanism again selects the high gain receiver, thus enabling continuous sensing operation.  To achieve minimal power consumption, the  passive receiver controls the switchover mechanism. 

\begin{figure}[!tb]
    \center
\includegraphics[width=0.4\linewidth]{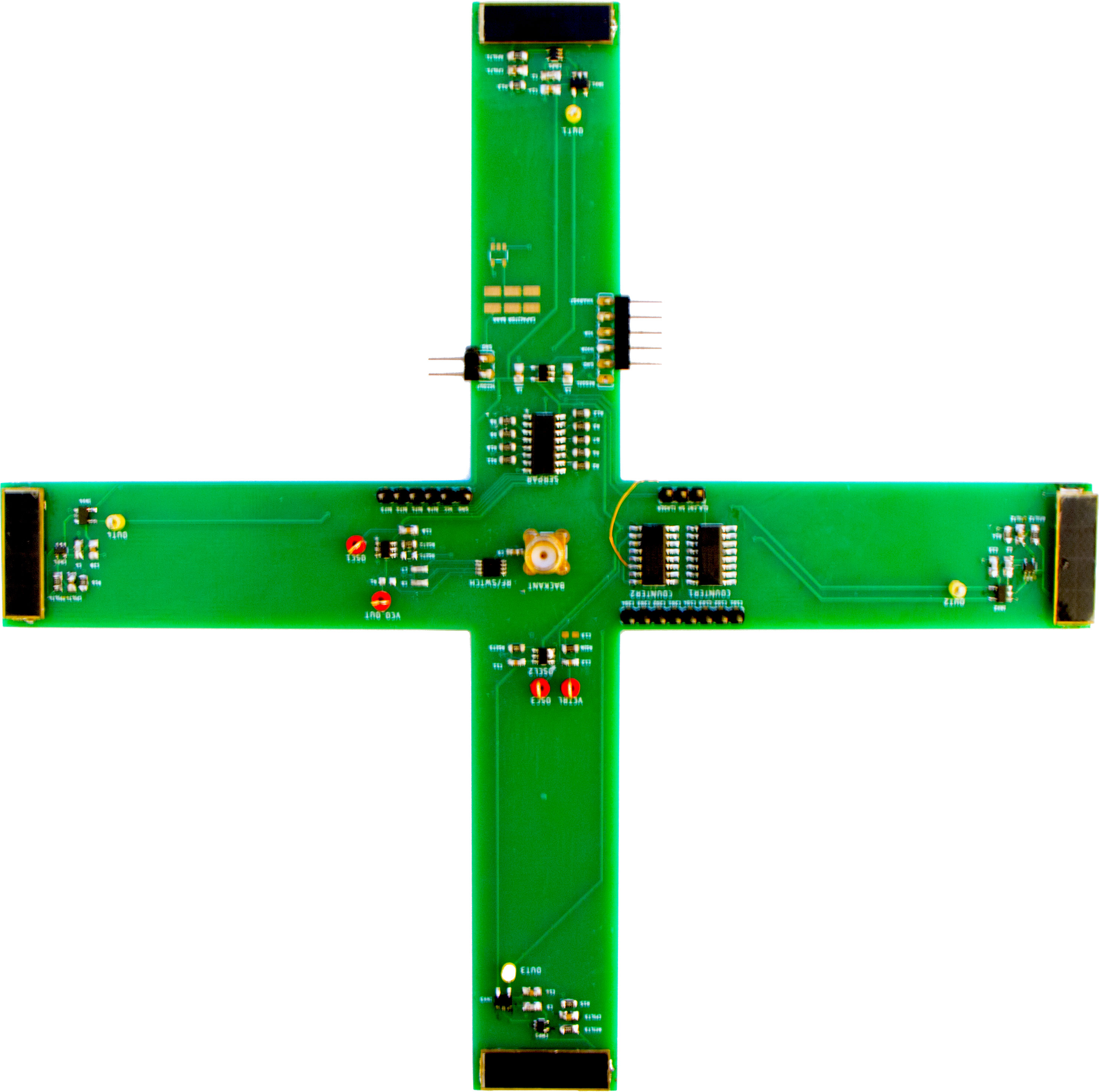} 
\vspace{-4mm}
\caption{\emph{Hand gesture sensing Prototype.} We detect hand gestures and transmit them with TunnelScatter.}%
    \label{fig:passiveswitch}%
    \vspace{-6mm}
\end{figure}

\fakepar{Hand gesture sensing}  In addition to the system described earlier, we design a version of the TunnelTag that we optimise for the lowest power consumption. We show the hardware prototype in Figure~\ref{fig:passiveswitch}. In this design, we employ several solar cells to detect the direction of hand movements, building on Li et al.~\cite{selfpoweredgest}. To achieve this we design the board with four passive receivers whose output is fused together using a parallel to serial converter before being transmitted using TunnelScatter. In the design, we also devise the transmission of preamble bits to enable support for several TunnelTags.

\begin{figure}[!t]
    \centering
    \includegraphics[width=0.47\textwidth]{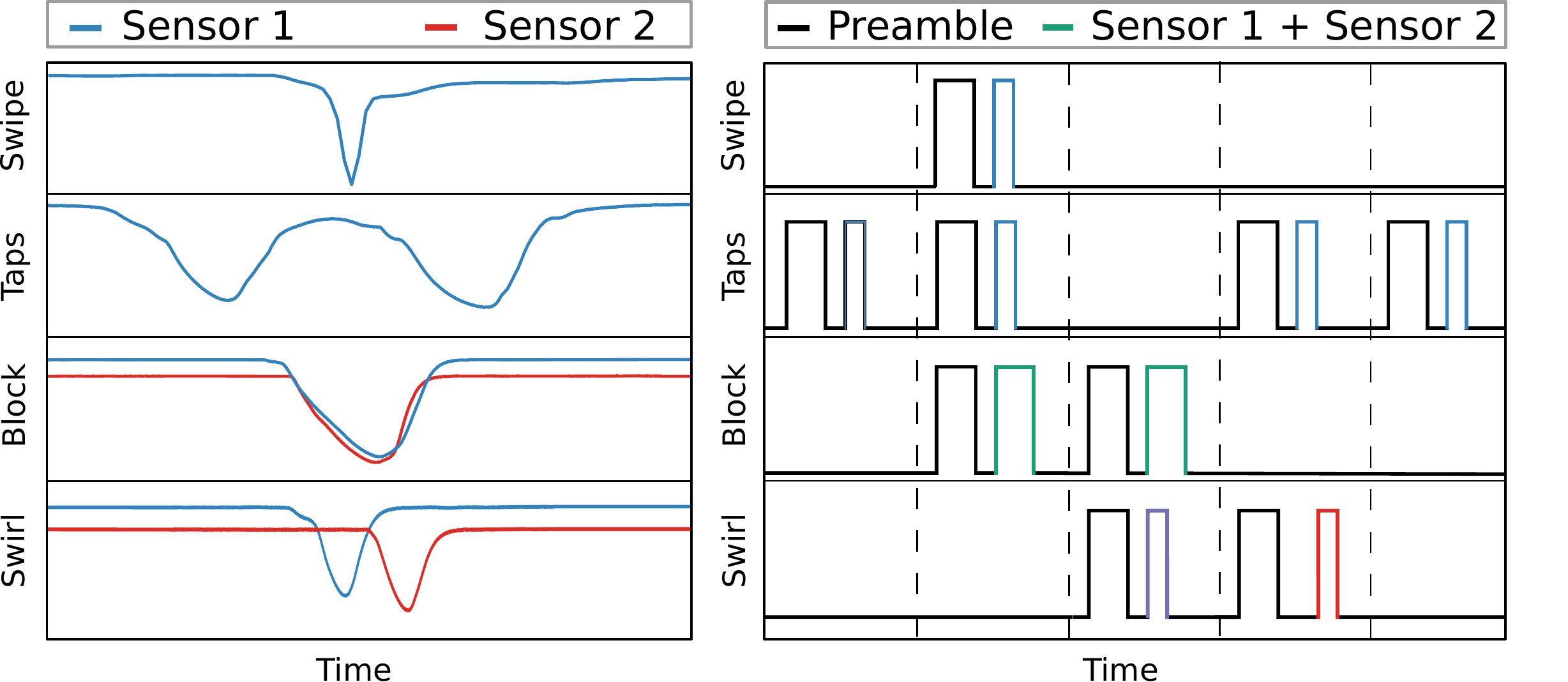} 
    \vspace{-4mm}
    \caption{\textit{Hand gesture sensing.} Through the optimised TunnelTag, each sensor is allocated a binary sequence, which appears after the preamble bits, as shown by the color coded digital signals. When two sensors are blocked, it gives rise to a new sequence.}
   \label{fig:gestureDetection}
    \vspace{-6mm}
    \end{figure}

\begin{figure}[!tb]
\center
    \includegraphics[width=0.9\linewidth]{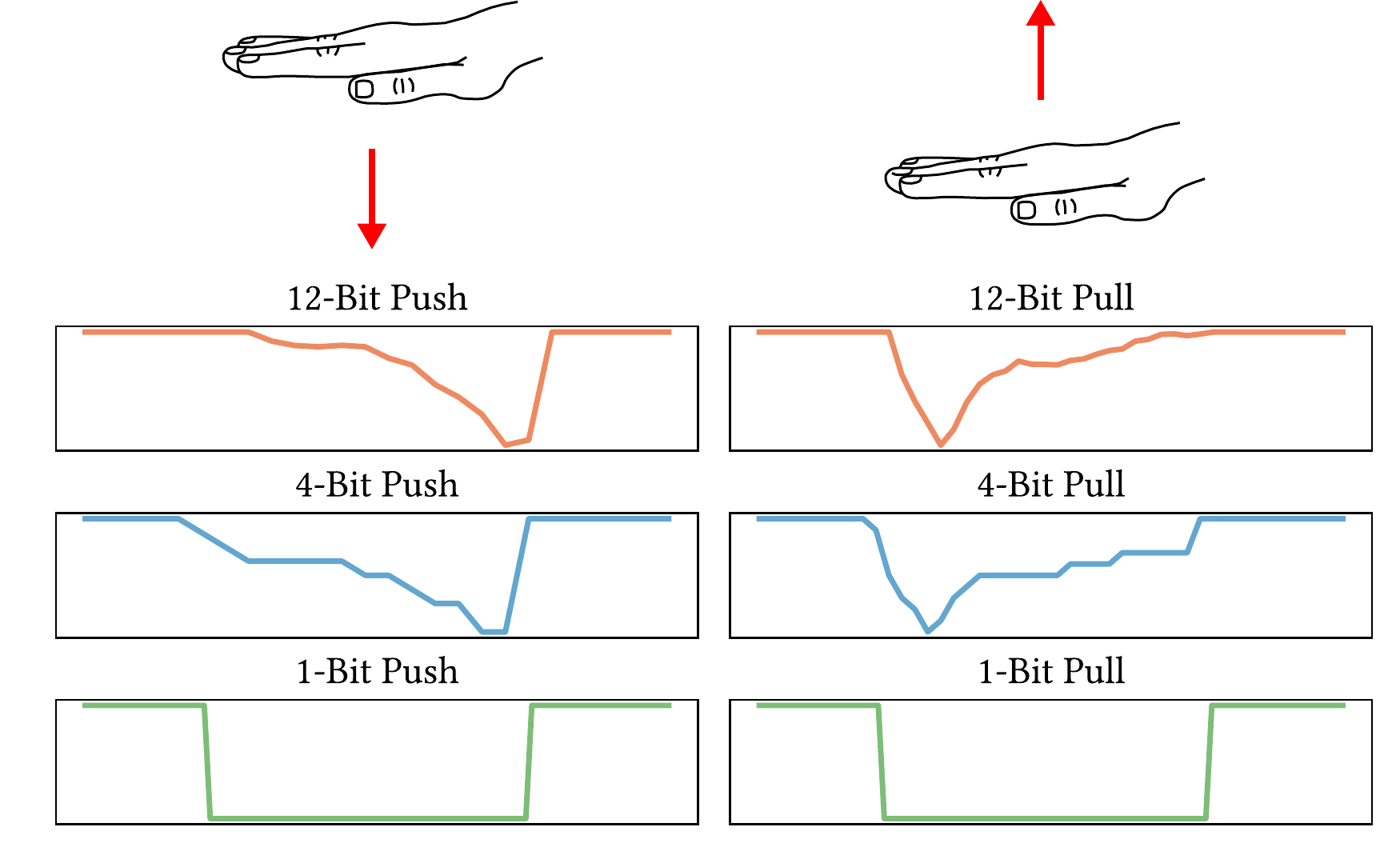}
        \vspace{-4mm}
    \caption{\textit{Multibit Gestures.} Complex gestures cause slow changes in the signal. Differentiating between the two requires higher sensing resolution than 1 bit.}
    \label{fig:gesture_pushpull} 
    \vspace{-6mm}
\end{figure}

With the optimised TunnelTag, we only make use of the LRP component to support four different hand gestures. We demonstrate the following gestures: swipe, taps, block, and swirl. A Swipe is represented as a brief finger movement in any direction over a single light sensor, taps is represented by lifting the finger above and away from a sensor a given number of times, and block is represented as a complete obstruction of the light sensor for a longer duration of time. Finally, a swirl is represented by moving a finger from one sensor to another in either a clockwise or a counterclockwise motion. Each of these gestures give rise to a unique pattern in the output of the passive receiver, as show in the Figure~\ref{fig:gestureDetection}. The gesture information is later transmitted, and then interpreted by the edge device. Next, we demonstrate how we can utilise the PPP, to enable more complex hand gestures which are not possible to recognize with the 1 bit design~\cite{bfvls} . We demonstrate two types of complex gestures, namely push and pull. A push is represented by a decreasing palm movement towards the sensor, and a pull is the reverse motion. In Figure~\ref{fig:gesture_pushpull}, we show the 4 and 12 bit representations of the sensed gestures.  The figure also shows that the 1 bit resolution produces similar signal for the Push and Pull gestures and cannot differentiate between them.

\fakepar{Related works}  A complimentary and concurrent effort to ours~\cite{solargest} develops techniques to detect hand gestures, which can be used with our system to support the detection of gestures with high accuracy. We focus on hardware innovations, such as low-power transmissions using TunnelScatter.

This application scenario is related to our previous design~\cite{bfvls}. 
However, we significantly improve this design in several ways. 
First, we improve the ability to operate in diverse light conditions through the self-sustaining light sensor. Next, 
we support multibit digitisation using PPP, which supports sensing and communication of complex gestures. Finally, we support transmissions in diverse ACS conditions through the TunnelScatter mechanism. Our system is also related to BackVLC~\cite{backvlc}. However, 
we significantly enhance its wireless communication ability through TunnelScatter.

\subsection{Other Scenarios}
\label{otherapplications}

\fakepar{Smart Homes} Recently there is an interest to develop various home automation systems. TunnelScatter could make many 
sensors used in this application battery-free, such as temperature or occupancy sensors. In a concrete example, there is an interest in designing wireless buttons~\cite{flic} which can be configured for different tasks, such as controlling appliances like television.  However, at present they use traditional radios such as BLE~\cite{flic}. We can improve the energy efficiency of these designs by designing a self-sustaining and tactile sensor, and integrating it with the TunnelTag.  

\fakepar{RFID Systems} RFID systems are widely used to support   applications such as inventory tracking, ID tags~\cite{want2006introduction}. RFID systems, however, are constrained due to power and cost limitations of the reader device. Systems like LoRea~\cite{lorea} overcome this by leveraging bistatic configurations, and using low-cost transceivers. However, such scenarios are  constrained as they require a powerful ACS emitter~(few \SI{}{\watt}s). TunnelScatter can improve such scenarios by eliminating the need for power hungry ACS emitters.

\fakepar{Visible Light Communication} VLC is an emerging communication medium for battery free sensor tags. Recent efforts support downlink  using VLC on battery free sensor tags~\cite{retrovlc,backvlc}. Energy efficient uplink communication, however, remains a challenge. Efforts like BackVLC~\cite{backvlc} address this by integrating a carrier generator in LED bulbs to support the uplink through backscatter. However, this is challenging  for practical deployments as it requires retrofitting LED bulbs with additional hardware.  TunnelScatter could simplify such scenarios by supporting uplink communication without requiring ACS, and  modifications to existing infrastructure. 

